\documentstyle[12pt,aaspp4]{article}

\newcommand{\ts}{\thinspace}
\newcommand{\p}{$^{\prime}$\ts}

\newcommand{\degree}{\arcdeg}
\newcommand{\msun}{\hbox{M$_\odot$}\ }

\newcommand{\lsunns}{\hbox{{\it L}$_\odot$}}

\newcommand{\lstar}{L$^*$}

\lefthead{Surace et al.}
\righthead{IR-Excess QSOs }

\begin{document}

\title{Optical/Near-Infrared Imaging of Infrared-Excess Palomar-Green QSOs}

\author{Jason A. Surace}
\affil{SIRTF Science Center, MS 220-6, California Institute of Technology, 
Jet Propulsion Laboratory, Pasadena, CA 91125\\
Electronic mail: jason@ipac.caltech.edu \\} 
\author{D. B. Sanders}
\affil{University of Hawaii, Institute for Astronomy, 2680 Woodlawn Dr., 
Honolulu, HI, 96822 \\
and Max-Planck Institute feuer Extraterestische Physik, Garching, Germany \\
Electronic mail: sanders@ifa.hawaii.edu\\
and\\}
\author{A. S. Evans}
\affil{Department of Physics, Math, and Astronomy, SUNY, Stonybrook, 
NY 11794-3800\\
Electronic mail: aevans@mail.astro.sunysb.edu}

\vspace{0.15in}
\centerline{\it To appear in the December, 2001 Astronomical Journal}

\begin{abstract}

Ground-based high spatial-resolution (FWHM $<$ 0.3--0.8\arcsec) optical
and near-infrared imaging (0.4---2.2 $\mu$m) is presented for a
complete sample of optically selected  Palomar-Green QSOs with far-infrared
excesses at least as great as those of ``warm'' AGN-like ultraluminous infrared
galaxies (L$_{ir}$/L$_{big-blue-bump} > 0.46$). In all cases, the host
galaxies of the QSOs were detected and most have discernable two-dimensional
structure. The QSO host galaxies and the QSO nuclei are similar in 
magnitude at H-band. H-band luminosities 
of the hosts range from 0.5-7.5 {\it L}$^*$ with a mean of 2.3 {\it L}$^*$, and are 
consistent with those found in ULIGs. Both the QSO nuclei and the 
host galaxies have near-infrared excesses, which may be the result of 
dust associated with the nucleus and of recent dusty star formation 
in the host. These results 
suggest that some, but not all, optically-selected QSOs may have evolved from an 
infrared-active state triggered by the merger of two similarly-sized L$^*$ galaxies, 
in a manner similar to that of the ultraluminous infrared galaxies.

\end{abstract}


\section{Introduction}

Many galaxies (e.g., Seyfert galaxies, radio galaxies)  have forms of 
abnormally energetic behavior lumped under the general heading of active 
galactic nuclei (AGN). IRAS discovered a new class of these AGN: galaxies
that had QSO-like bolometric luminosities ({\it L}$_{\rm bol} >$ 10$^{12}$ {\it 
L}$_{\sun}$) yet which emit nearly all of this
luminosity at far-infrared wavelengths. Nearly all of these galaxies show evidence 
for merger activity along with optical line ratios similar to that of other AGN 
(Sanders et al. 1988a, Murphy et al. 1996, Veilleux et al. 1999). Tying together 
the theoretical framework for merger-induced fueling (Toomre \& Toomre 
1972, Barnes \& Hernquist 1996) and the known merger structure around some 
QSOs (Stockton \& MacKenty 1987, Hutchings et al. 1995) along with an 
evolutionary timeline, Sanders et al. (1988b) proposed that these so-called 
ultraluminous infrared galaxies (ULIGs) were the immediate progenitors of 
optically selected QSOs, and that ULIGs represented an evolutionary 
stage 
where the QSO nuclei were enshrouded in gas and dust. This dust shroud 
reradiates the nuclear emission at far-infrared wavelengths; eventually, 
powerful winds (Heckman et al. 1990) disperse the dust and reveal an optical 
QSO nucleus.

Surace \& Sanders (1998, 1999, 2000b; hereafter called Papers I, II, and IV), 
Surace et al. (2000a; hereafter called Paper III), and Scoville et al. (2000) 
explored the morphology and colors of the circumnuclear and extended 
environments of ULIGs at optical and near-infrared wavelengths. They 
demonstrated that the emission at optical and near-ultraviolet wavelengths 
from ULIGs 
is dominated by an extended starburst many kpc in size. A compact 
core, however, dominates at near-infrared wavelengths. This result was 
dramatically confirmed at mid-IR wavelengths by Soifer et al. (2000) 
and thus directly demonstrated that the bolometric luminosity arises 
in very small region. If QSOs are relic 
ULIGs, then their host galaxies should provide evidence for the aged remnants 
of the structures found in ULIGs: faded extended tidal tails and debris, and aged star-
forming knots from a tidally-triggered starburst.

At the same time, considerable advances were made in imaging studies of 
QSO host galaxies. Ground-based studies using high resolution techniques 
clearly revealed the host galaxies of many QSOs (Dunlop et al. 1993, McLeod \& 
Rieke 1994ab) and in some cases the observed morphologies could be 
attributed to interactions (e.g. Hutchings \& Neff 1992). More recent results have 
been obtained using the high spatial resolution of {\it HST}. McLure et al. (1999) have shown 
that the 
majority of high luminosity quasars (both radio-loud and radio-quiet) lie in elliptical hosts, while lower 
luminosity systems (primarily radio-quiet) lie in a wider variety of hosts. 
This is in contrast to earlier 
studies, which had difficulties detecting the hosts of QSOs (e.g. Bahcall 
et al. 1995).
 
By observing a complete sample of QSOs with bolometric and infrared 
luminosities similar to those of 
ULIGs and using the same instruments and techniques as Papers I---IV, the 
question of the relationship between the structure of the QSO host 
galaxies and ULIGs can be investigated. The development of high quantum 
efficiency, large format detectors spanning the wavelength range from the 
near-ultraviolet (3000\AA) to the near-infrared (2.2\micron) enable detailed 
studies over a wide range of wavelengths dominated by different emission 
processes. The advent of adaptive optics techniques coupled with superior 
ground-based sites also enable imaging with much 
higher spatial resolution than ever before possible, simplifying detection of QSO 
host galaxies and any compact features within them.

A general feature of the ULIG---QSO evolutionary scenario is that the spectral 
energy distributions (SEDs) must evolve from a state dominated by far-
infrared emission (ULIGs) to a state where optical/ultraviolet emission 
becomes a strong, if not dominant component of the SED. The previous studies 
of ULIGs by Surace et al. (Papers I-IV) concentrated on samples of ``cool'' ULIGs 
($f_{25}/f_{60} < 0.2$)\footnote{The quantities $f_{12}$,
$f_{25}$, $f_{60}$, and $f_{100}$ represent the {\it IRAS} flux densities in Jy
at 12{\ts}\micron, 25{\ts}\micron, 60{\ts}\micron, and 100{\ts}\micron\
respectively.}, 
the least evolved with the most far-infrared dominated 
SEDs, and ``warm'' ULIGs ($f_{25}/f_{60} > 0.2$) which have a stronger 
relative optical/ultraviolet component.  The ``youngest'' QSOs (i.e. those closest 
to the ULIG evolutionary state) must  be those that have the greatest far-
infrared component to their SEDs. Selection of such ``infrared-excess'' QSOs  
should therefore select the most ULIG-like systems.

We report the results of an imaging survey at B, I, H, and K\p of
a complete sample of low redshift ({\it z}$<$ 0.16) QSOs selected from the 
Palomar-Green Bright Quasar Survey (Schmidt \& Green 1983) on the basis of 
having an infrared-excess greater than the least IR-excess ``warm'' ULIGs and which 
therefore, within the context of the evolutionary scenario,
should be the youngest examples of QSOs. Their morphologies, 
luminosities, and colors are compared 
to those of the ULIGs (Papers I---III), and the implications for the ULIG-QSO 
evolutionary scenario are discussed.

\section{Data}

\subsection{Sample Selection}

The sample is drawn from all Palomar-Green ( PG; 
Schmidt \& Green 1983) Bright Quasar Sample QSOs with redshifts {\it z}$\leq$0.16, thus placing 
them in the same space volume as the ``cool'' and ``warm'' ULIG samples 
examined in Papers I \& III. This is important because the interpretation of 
galaxy morphology often depends on the achieved physical spatial resolution. 
For example, the large ``knots'' in NGC 4038/9 (The Antennae Galaxy) seen from the ground are 
revealed by HST to be composed of smaller subclumps of star forming 
clusters (Whitmore \& Schwiezer 1995).  
By examining a sample with the same techniques used in Paper III and 
at the same distance, the 
morphologies of the host galaxies are likely to be interpreted in a similar fashion. 
Additionally, the 
low redshift of the sample has the obvious advantage that as the closest examples of 
their population they are the most amenable to detailed study.

As described above, in order to find transition objects which bridge the gap between 
``warm'' ULIGs 
and the general QSO population an infrared selection criterion was imposed. 
The ratio L$_{ir}$/L$_{BBB}$\footnote{$L_{\rm ir} 
\equiv L$(8---1000$\micron)$  was computed using the flux in all four {\it IRAS} 
bands according to the prescription given in Perault (1987). {\it L}$_{\rm 
BBB}$ was computed from the ``Big Blue Bump'' luminosity at 3200--8400\AA \ 
and the photometry of Sanders et al. 
(1989)} was used to find QSOs with 
``excess'' far-IR emission.
The quantities L$_{ir}$ and L$_{BBB}$ were computed from the data given by Sanders et 
al. (1989). The warm ULIG with the lowest such value is 3C273, which has 
L$_{ir}$/L$_{BBB} =$ 0.46.
Only PG QSOs with L$_{ir}$/L$_{BBB} >$ 
0.46 were selected. 
The distributions of L$_{ir}$/L$_{BBB}$ for the QSO sample 
and the ULIG samples are not the same, nor could they be, given that ULIGs by definition 
almost always have very high L$_{ir}$/L$_{BBB}$. Instead, they have 
the same lower bound. This criterion 
was only used to search for ``red'' QSOs which might have significant dust 
content and hence be more evolved versions of ULIGs. All three 
samples {\it do} have the same redshift distribution (Paper III).

Ideally, the QSOs would also have the same distribution of bolometric luminosities as 
ULIGs since it is hypothesized that the infrared luminosity in ULIGs is QSO-
driven. In practice this is difficult to achieve. Besides being well-
studied, the PG QSO sample was the only one at the time of this project with 
well-characterized far-IR emission.
The PG BQS has too few objects meeting both our redshift and infrared color 
criterion to allow the construction of a meaningfully large sample with the 
same bolometric luminosity distribution as the ULIGs. Future studies 
could avoid this problem by using other infrared QSO surveys, such as those 
undertaken by ISO (Hooper et al. 1999) or SIRTF. Therefore, the QSO sample was 
selected instead to meet the same minimum bolometric luminosity criterion for 
ULIGs (L$_{bol} > 10^{12}$\lsunns ). The bolometric luminosity was 
determined in two ways. First, L$_{bol}$ was estimated from L$_{bol}$ = 16.5 
$\times \nu L_{B}$ where L$_B$ is the luminosity at B-band, which is the 
average bolometric correction found for PG QSOs by Sanders et al. 
(1989), although more recent estimates of this correction factor 
indicate a value of 11.7. 
Under the former definition the luminosity L$_{bol}$=10$^{12}$ \lsunns corresponds 
to M$_B$=-22.18. Papers identifying ULIGs have principally used a different 
cosmology than Schmidt \& Green (e.g. H$_0$=75 km s$^{-1}$ Mpc $^{-1}$ vs. 50 
km s$^{-1}$ Mpc $^{-1}$), adjusting for this implies that QSOs have M$_B <$ -23, 
similar to the definition found in that paper. Second, where possible 
L$_{bol}$ 
was determined directly from the data of Sanders et al. (1989). Since L$_{bol}$ 
for UGC 5101 is slightly below 10$^{12}$\lsunns, the range in bolometric 
luminosities was allowed to drop to Log L$_{bol}$=11.92. Two objects in our 
sample (PG 1126-041 and PG 1229+204) were included because of this.

There are 17 infrared-excess PG QSOs, out of a total of 36 candidates meeting the 
redshift and bolometric luminosity criterion. However, 3 of these (I Zw 1, Mrk 
1014, and 3c273) are also ``warm'' ULIGs and have already been discussed in 
detail in Papers I and II. One additional object was observed (PG 1119+120). 
Although originally in our sample, a re-evaluation of its bolometric 
luminosity showed that it fell substantially short of our luminosity criterion, even though it has 
historically been considered a QSO (Neugebauer \& Matthews 1999, Rowan-
Robinson 1995). The data for this object are presented, but not used 
in the following discussion. Table 1 presents the complete sample 
list, along with measured values of L$_{ir}$/L$_{BBB}$ and  L$_{bol}$. 
Note that the table presents new values for these quantities derived 
from the new photometry presented in this paper and our current 
understanding of the relationship between bolometric luminosity and 
the luminosity at specific wavelengths, not the photometry 
originally
used to select the sample.

\subsection{Observations and Reduction}

The optical data were taken at the f/31 focus of the University of Hawaii 2.2m 
telescope using a fast tip/tilt guider in the manner described 
in Papers II \& III. B \& I-band data were acquired using the 
Tektronix 2048 and Orbit 2048 cameras between May 1996 and December 1997. 
In most cases this was by direct imaging at the f/31 Cassegrain focus. The data 
were binned on chip 2x2, with pixel scales of 0.14 
and 0.09\arcsec , respectively. In a few cases the data were obtained by 
reimaging the f/31 beam at f/10 through the HARIS spectrograph using the 
Orbit 2048 camera with a pixel scale of 0.14\arcsec . While the fast tip/tilt 
guider does not appreciably correct atmospheric seeing at optical 
wavelengths, it can eliminate common-mode vibrations such as wind shake 
which contribute appreciably to image degradation. Typical spatial resolutions 
varied between 0.4---0.9\arcsec at I-band, and with B-band being 
substantially worse by as much as factor of 2. Total exposure times varied between 12 and 
54 minutes per filter per target. Individual exposure times were short enough 
to prevent saturation 
of the QSO nucleus. In some cases the Cassegrain focus was rotated in order to 
improve offset guidestar acquisition, in which case additional flat-field images 
were taken at the appropriate rotations.

The optical data reduction involved several steps. First, the CCD bias pattern 
was removed by subtracting from each image a high S/N median bias frame 
constructed from sequences of 20-30 bias frames at the beginning and end of 
each night. Pixel-to-pixel response variations were corrected by dividing 
each image by a high S/N flat produced by making dithered observations of 
the twilight sky in each filter. Typical twilight exposures were 2-3 seconds 
each; short enough to avoid getting detectable flux from field stars, yet long 
enough to avoid flat-field errors introduced by the radial shutter used at the 
UH 2.2m. Estimated S/N for the flats (based on Poisson statistics and the gain of 
the CCD) was between 250-500. Neither CCD showed any evidence of measurable 
dark current, based on an examination of long closed shutter exposures. The 
images were corrected to normal orientation by transposition and 
rotation using the ROTATE task in IRAF based on the known field rotation of 
the Cassegrain focus of the UH 2.2m, which is accurate to better than 1 degree. 
The CCD overscan regions were trimmed using IMCOPY. The images were then 
aligned using IMALIGN task, which uses a marginal centroiding 
routine that calculates a best fit solution to a number of (user-supplied) 
reference stars in the field. Typical alignment errors were estimated (on the 
basis of the fit) to be about 0.25 pixels. The data were typically 
sampled at 5 pixels FWHM for a point source so alignment errors are unlikely to 
be important. The images were then averaged using an algorithm that rejects 
pixels inconsistent with the known noise properties of the CCD thus 
rejecting cosmic rays. The shifted images were combined onto a larger 
image than the original data frames thereby increasing the total field of view 
due to the dithering process. This was valuable in order to increase 
the availability of PSF stars since the camera field of view was much larger 
than the measurable extent of any of the galaxies.

The near-infrared H \& K\p data were acquired between August 1996 and 
January 1998 with the QUIRC 1024$^2$ HgCdTe camera (Hodapp et al. 1996) at 
the UH 2.2m using the f/31 Cassegrain focus and the tip/tilt guider. At this 
focus the plate scale is 0.06 \arcsec pixel$^{-1}$. The K\p filter was chosen 
due to its lower thermal background and hence greater sensitivity (Wainscoat 
\& Cowie 1992). Throughout this paper we exclusively refer to K\p. Comparison 
to work by other authors is made using the conversion of Wainscoat \& Cowie 
(1992) from UH K\p to CIT K. In all cases exposure times were short enough 
(60--240 seconds) to prevent saturation of the QSO nucleus, thus allowing PSF 
subtraction during post-processing. Typical achieved spatial resolution using 
tip/tilt was 0.25---0.5\arcsec . Total exposure times varied from 9 to 45 
minutes per filter. Details of the new observations presented in this paper are 
given in Table 1. Readers should note that the tabulated values of 
L$_{ir}$/L$_{BBB}$ reflect updated values based on the new photometry 
presented here, and not those used for the initial sample selection. 
These new values reflect changes primarily in the optical properties 
of the QSOs and also a refined model for the QSO SED, and mostly affect 
PG 0007+106 and PG 1351+640.

The near-infrared data were reduced in the same manner as that described in 
Paper II. The data was initially sky-subtracted using consecutive, dithered 
frames; because the QUIRC field of view is so large (32\arcsec ), it was possible 
to dither the target on-chip, thereby increasing telescope efficiency by a 
factor of 2. Responsivity variations were removed using median flats constructed 
from images of the illuminated dome interior. Each image was masked by 
hand to exclude bad pixels and regions contaminated by negative emission 
introduced by the sky subtraction. Images were aligned using the method 
described above for the optical data. Images were scaled according to their 
exposure times and then, in order to account for any variable sky background, 
an offset was subtracted from each image based on the background actually 
measured in that frame. The images were then combined by medianing using 
IMCOMBINE and rejecting pixels outside the linear repsonse of the array.

One QSO, PG{\ts}1202+281, could not be observed as there were no guide stars 
sufficiently bright ({\it m}$_{\rm V} < $ 13) and nearby ( {\it radial 
distance} $<$ 
5\arcmin ) to enable the use of the tip/tilt guider. This QSO was, however, 
observed by {\it HST}/WFPC2.
Several other QSOs (PG 1229+204, PG 1402+261, and PG 2130+099) have also been 
observed with {\it HST}/WFPC2 as part of other observing programs 
(Hutchings \& Morris 1995, Bahcall et al. 1997), and were available 
from 
the STScI data archive. The WFPC2 images were taken 
through the F606W  and F702W filters, which correspond roughly to V and R-
band and are therefore not directly comparable to our data. They are, 
however, valuable for interpreting morphology that is only marginally 
resolved from the ground. This data was reduced in the same manner as described by Surace et al. 
(1998); i.e., by shifting and rotating according to the astrometric solution 
provided by STScI, and combined using the GCOMBINE task in IRAF/STSDAS, 
with cosmic ray-rejection. Residual glitches were interpolated by hand using 
IMEDIT.
The three QSOs which are also warm ULIGs (I Zw 1, Mrk 1014, and 3c273) 
have also been observed with {\it HST} (Paper I, Bahcall et al. 1995); 
since these 
data are similarly analyzed elsewhere in the literature, they are not presented here.

The ground-based data were flux calibrated using observations of 
Landolt (1983, 1992) optical standards and 
Elias (1982) infrared standard stars. In most cases the nights were photometric with $\sigma_{\rm M} 
<$ 0.05. For data taken on non-photometric nights, the data were 
calibrated
using large fixed-aperture photometry already in the literature for the 
targets. Specifically, the near-infrared data for PG 0007+106 and PG 1001+054 
were calibrated using the photometry given by Neugebauer et al. (1987) and the 
magnitude zeropoints contained therein. Optical data for PG 1119+120, PG 
1440+356, PG 1613+290, and PG 2130+099 were calibrated using the photometry 
of Neugebauer et al. (1987) and the magnitude zeropoints of Bessel (1979). 
Photometric calibration errors are 0.03 magnitudes. 

The point-spread-function (PSF) was calibrated from stars in the 
final combined science images using DAOPHOT as described in 
Paper II. The stars were identified, scaled, shifted, and combined using a 
sigma-clipping algorithm and weighted according to total stellar flux, thus creating 
as high a S/N PSF image as possible. In those few cases where no stars were 
found in the science images, the PSF was estimated by using the closest 
temporally adjacent PSF. Since the tip/tilt guiding has little effect on 
atmospheric distortions at short wavelengths, this works well for optical data. 
Similarly, since the seeing remains stable on timescales of many minutes, this 
is also effective in the near-infrared.

The integrated photometry was derived by measuring the total flux of the 
QSO/host system in an aperture large enough to encompass the optical extent 
of the galaxy at a flux level below 1 $\sigma$. Measurement errors from this 
technique are small and are dominated by the photometric calibration errors. 
The host galaxy luminosity was then derived by subtracting the contribution 
of the QSO nucleus and any high surface brightness features (e.g., star 
forming knots). As in Paper II, the luminosity of the QSO nucleus was 
determined in two ways: measuring the flux in a fixed 2.5 kpc radius aperture 
and then correcting this with an aperture correction derived from the 
observed PSF, and also by subtracting the observed PSF scaled to the 
nuclear luminosity such that a 
smooth host galaxy flux distribution without holes resulted. Generally, these 
results were the same to within 10\%. This is not surprising since the surface 
brightness of the QSO nucleus is much higher than that of the underlying 
galaxy and small aperture photometry will mostly be sensitive to the QSO nucleus 
since the underlying host contributes only a small fraction of the total flux inside 
the aperture. The uncertainties in the host and nuclear luminosities 
are 
dominated by the measurement process and are 0.12 magnitudes.

Isophotal profiles were determined using the IRAF.STSDAS routine ELLIPSE. This 
routine builds elliptical isophotal models using an iterative 
technique. For the radial profiles the center and position angle were 
held fixed. For a discussion of how these profiles were computed when 
searching for bars, see \S 3.1.1.

\section{Results}

\subsection{Morphology}

\subsubsection{Large-Scale Features}

Host galaxies (as evidenced by extended emission beyond that expected for a 
point source) were detected around every QSO observed. Figure 1 presents the large-scale 
morphology of the IR-excess
QSOs (except I Zw 1, Mrk 1014, PG 1202+281, and 3c273) at all four observed 
wavelengths. In each case the images have been stretched in an attempt to 
emphasize the faint structure in the host galaxies. Figure 2 presents  near-truecolor 
images for every QSO including I Zw 1 and Mrk 1014 for completeness but 
excluding PG 1202+281 and 3c273 (we have no multicolor optical 
data for these two objects). These images are the most intuitive to understand 
and hence the most instructive for the following discussion (Surace et al. 
2000a). Figure 3 presents {\it HST} data for several objects in our 
sample, and is useful for comparison with Figure 1.

\begin{figure}
\vspace{2.5in}
\centerline{\bf These figures are provided as JPGs.}
\vspace{2.5in}
\caption{BIHK\p data for the infrared-excess PG QSO sample. Image 
intensities have a log scaling to accentuate the low surface brightness 
features in the host galaxies. The B and I-band data have been smoothed by 
convolution with a 0.36\arcsec gaussian kernel. Ticks are every 2\arcsec , 
with major ticks every 10\arcsec . The scale bar is 10 kpc. NE is at top 
left.}
\end{figure}

\begin{figure}[htbp]
\epsscale{0.7}
\plotone{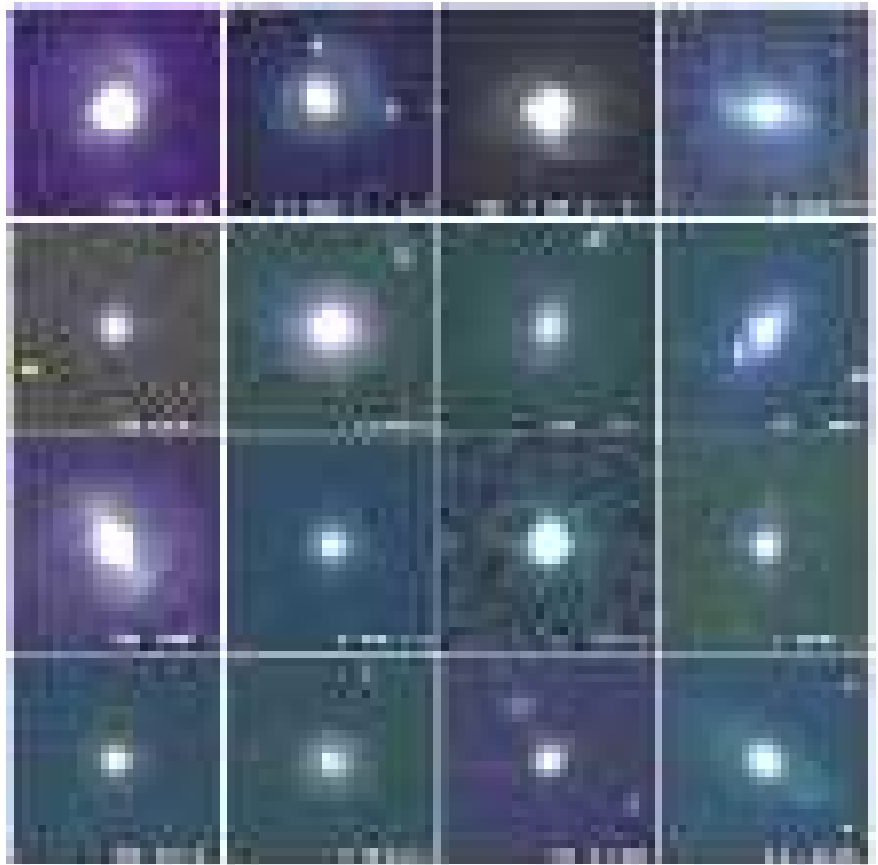}
\caption{Near-truecolor images of the QSO sample constructed from the B 
\& I-band data. The galaxy SEDs have been linearly interpolated from the {\it 
B} and {\it I} data; the color balance is not absolute. For completeness, I Zw 1 
and Mrk 1014 have been included from Paper I. 3C273 and PG 1202+281 do not 
appear since B \& I observations of these QSOs were not obtained.{\bf 
This figure is provided separately as a JPG.}}
\end{figure}

\begin{figure}[htbp]
\epsscale{0.6}
\plotone{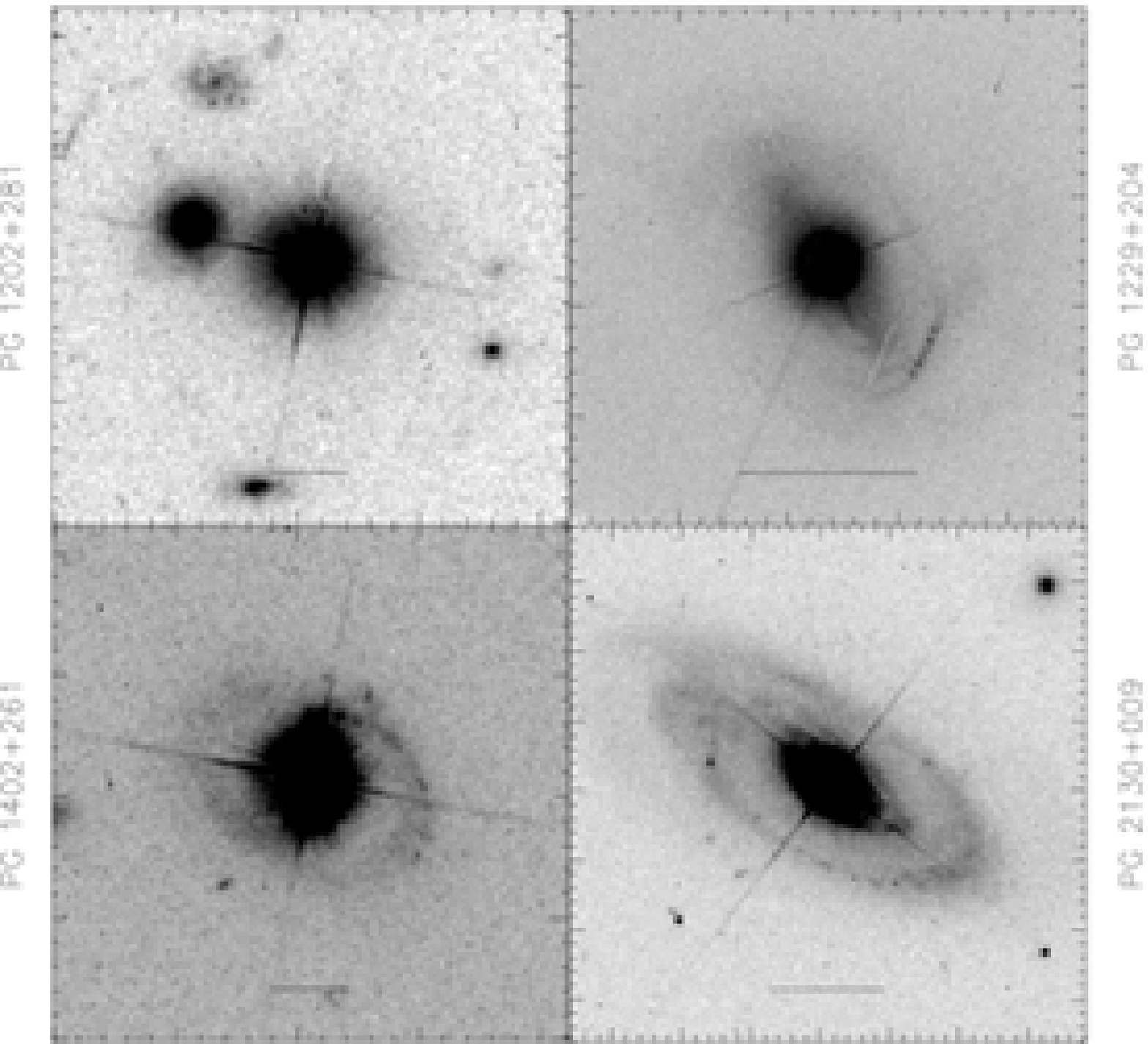}
\caption{Archival images of infrared-excess PG QSOs taken with {\it 
HST}/WFPC2 in the V and R filters as part of other programs (Bahcall et al. 
1997, Hutchings \& Morris 1995). Tick marks are 1\arcsec , and the physical 
scale bar is 10 kpc. {\bf This figure is provided separately as a JPG.}}
\end{figure}

Radial profiles are presented in Figure 4. These profiles are the mean surface 
brightness at a given radius at H-band. Many studies of galaxy types have 
employed radial profile fitting, but at optical wavelengths. This is problematic 
for examining QSO hosts due to contamination of the radial profile by the 
nuclear PSF. Therefore, we present radial profiles at H-band, where the spatial 
resolution is typically 3$\times$ higher than at B-band. Most of the galaxies 
are extended with radii of 8---15 kpc. 
No attempt is made to classify the host galaxies as strictly spiral or 
elliptical-like based solely on their radial profiles. 
The interpretation of radial profiles is fairly ambiguous 
when only the outer profile is well-defined (Surace \& Sanders 2000). The 
differences between the de Vaucoleurs profile and the exponential disk profile 
are most unambiguous at small radii. However, the surface brightness 
distribution at small radii is dominated by the QSO nucleus, and incomplete 
knowledge of the PSF prevents the recovery of sufficient information about 
the host at these radii. Also, the merger scenario that is examined 
here is believed to transform spiral galaxies into post-merger 
systems that strongly resemble ellipticals, and hence transition 
objects could appear similar to either type, or neither. Instead, we consider the 
operative definition that 
elliptical galaxies have predominantly smooth surface brightness 
distributions. High surface brightness, high spatial frequency non-radially 
symmetric structure (such as spiral arms) are the surest sign that a 
host 
is not an elliptical galaxy. Other approaches which derive models from the data 
which can then be classified neccessarily ignore the presence of 
complex two-
dimensional structure, which is the spatial component of greatest interest to 
us. It is unclear whether our profiles 
resemble more closely disks or a de Vaucoleur's profile. In any case, since H-
band is dominated by the old stellar population, it is likely that any spiral 
profiles are strongly influenced by the presence of central bulges, further 
blurring interpretation.

\begin{figure}[htbp]
\epsscale{0.65}
\plotone{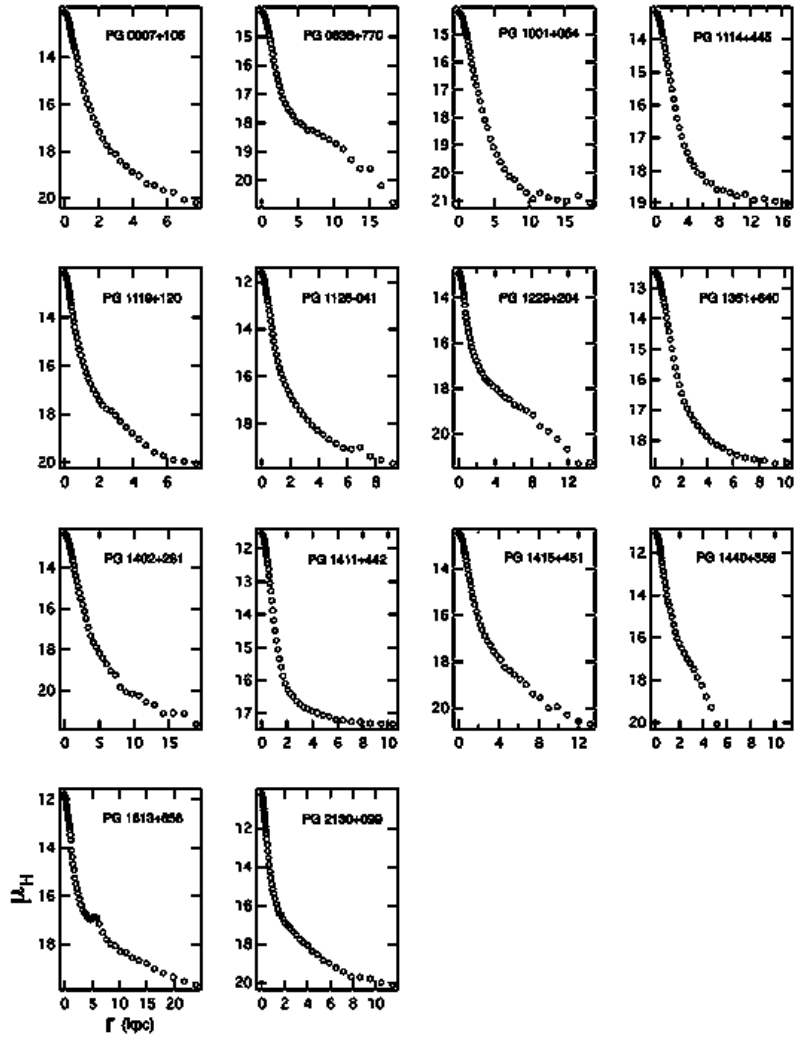}
\caption{H-band radial profiles of the observed QSOs. In this 
magnitude
representation an exponential disk profile will appear as a straight 
line.
{\bf This figure is provided separately as a JPG.}}
\end{figure}

Of the 17 QSOs, 8 are in spiral-like galaxies based on the presence of spiral 
arms or rings and bars. Four of these hosts have 
weak nuclear 
bars whose major axes have radii ranging from 5 to 11 kpc. The presence of bars (beyond 
qualitative morphology) was determined using the criteria of Knapen 
et al. (2000). A galaxy is considered barred if it has a significant rise in 
ellipticity, followed by a corresponding decrease, or a change in 
semi-major axis position angle of greater than 75\degree. Figure 5 
shows the radial profiles of the four QSO hosts, as well as PG 
1119+120, that meet these criteria 
and which do not have obvious tidal tails consistent with the galaxy 
being a major merger. The ellipticity 
of the inner isophotes relative to the nearly radial symmetric outer 
isophotes is most clearly seen in Figure 1. The bars have mean projected 
peak ellipticities of 0.36, where the ellipticity is defined as 

\begin{equation}
\epsilon=1-(b/a)
\end{equation}

\noindent where b and a are the semi-minor and semi-major axes, 
respectively. The deprojected mean peak ellipticity is 0.43.

\begin{figure}[htbp]
\epsscale{0.62}
\plotone{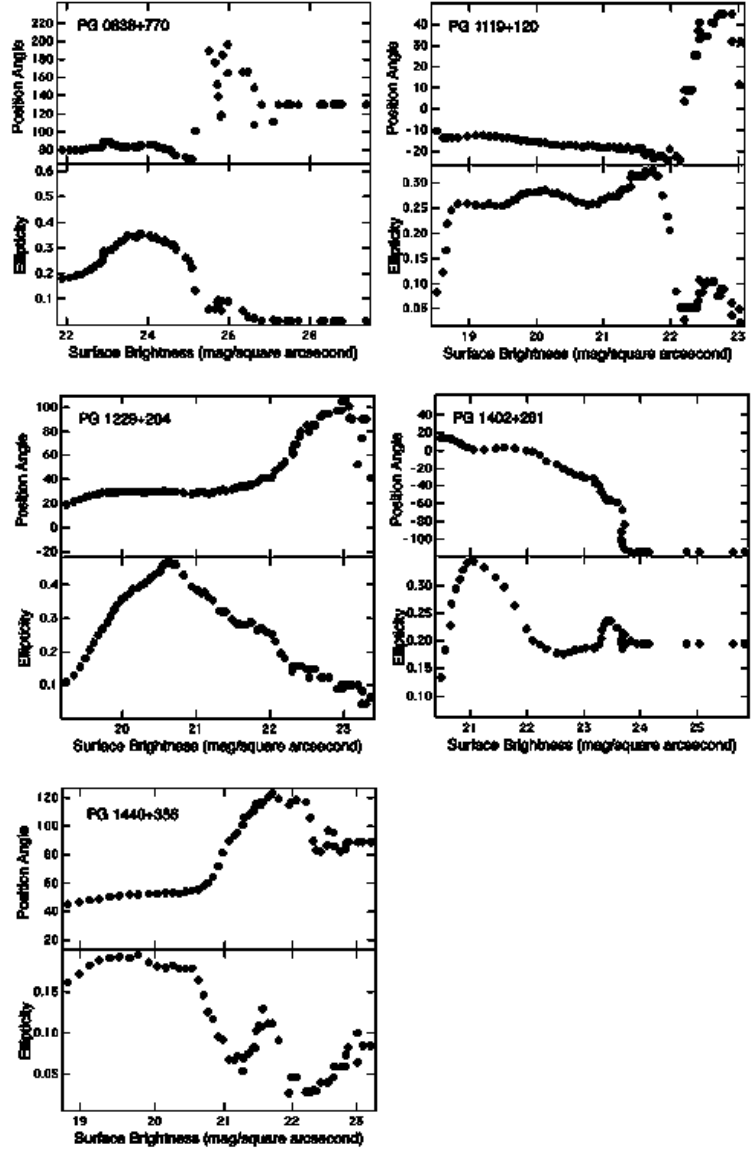}
\caption{I-band ellipticities and position angles as a function of 
isophotal magnitude. In these systems, the rise and subsequent fall in 
ellipticity is due to the presence of the bar at surface brightnesses 
intermediate between the point-like QSO nucleus and the more circular 
galaxy. In cases where the body of the galaxy is at a considerable 
projected angle, there is also a large change in position angle as the 
isophotes twist from the bar to the main galaxy body. {\bf 
This figure is provided separately as a JPG.}
}
\end{figure}

Two galaxies are known to be ellipticals: 
3c273 and PG 1202+281 (Bahcall et al. 1997). McLeod et al. (1994b) claim a 
detection for a host galaxy in the latter but give no information on spatial structure. The 
remainder (PG 1001+054, PG 1351+640, \& PG 1415+451) are indeterminate and 
lack any obvious structure in their extended emission. 
These three indeterminate cases also have very 
high nuclear luminosity fractions, thus making them hard to classify. 
An additional four host galaxies (PG 0007+106, PG 0157+001, PG 1411+442, and PG 
1613+658) have prominent tidal tails and arms and are 
unmistakably recent merger systems. Thus, at least 24\% (4/17) of the far-
infrared excess PG QSOs occur in merger systems.  Moreover, 
Hutchings et al. (1992,1994,1995) classified PG 1229+204 as a merger involving 
a small companion resulting in the observed bar and ring structures. 
Hutchings \& Neff (1992) also propose that PG 2130+099 may also be in a post-
interaction state, although it is not clear if it's disturbed morphology 
necessarily implies merger activity. Our 
morphologies generally agree well with existing published data (Hutchings \& 
Neff 1992, Dunlop et al. 1993, McLeod et al. 1994ab) .

\subsubsection{Star-Forming Knots}

Compact high surface brightness emission regions are detected in 35\% 
(6/17) 
of the QSO hosts. In particular, ``knots''\footnote{``knots'' are defined as closed 
isophotes less than 1 arcsecond in radius and 3 $\sigma$ above the 
surrounding emission.} of star formation like those found in the ULIGs 
(Surace et al. 1998,1999) are detected in PG0007+106, I Zw 1, Mrk 1014, PG 1229+204, PG 1411+442 and PG 1613+658.

Figure 6 shows the small-scale structure, enhanced by a 
variety of techniques. The optical data for I Zw 1 and Mrk 1014 appear in 
Figure 2, while the near-infrared data appear in Paper II. In most cases the observed PSF had 
insufficient S/N to be used for a high quality point source subtraction and so a 
noiseless model of the QSO nucleus was used instead. In Figure 6 radial profile 
models have been fit to the QSO nuclei using the ELLIPSE and BMODEL routines 
in IRAF/STSDAS. The models were forced to be circular, with centers fixed on 
the QSO nuclei. They were then subtracted from the raw data, leaving only the 
non-radially symmetric component of the QSO host galaxy behind. In several cases a 
modified version 
of this technique was used. Radial profile models were fit using the JIP 
imaging package in a manner similar to that above; only in this case the 
fitting region for the radial models was restricted to specific position angles, 
allowing exclusion of obvious structure such as stars. This results in a more 
accurate model which has less tendency to oversubtract from the host galaxy. 
Even with 
this technique, structure within a radius of 1--2\arcsec (2--4 kpc) of the 
nucleus generally cannot be recovered at optical wavelengths due to 
confusion with subtraction artifacts arising due to pixel aliasing. 
While the observed spatial resolution is generally mugh higher in the 
infrared, the host galaxies have 
much less structure at these wavelengths.

\begin{figure}
\epsscale{0.65}
\plotone{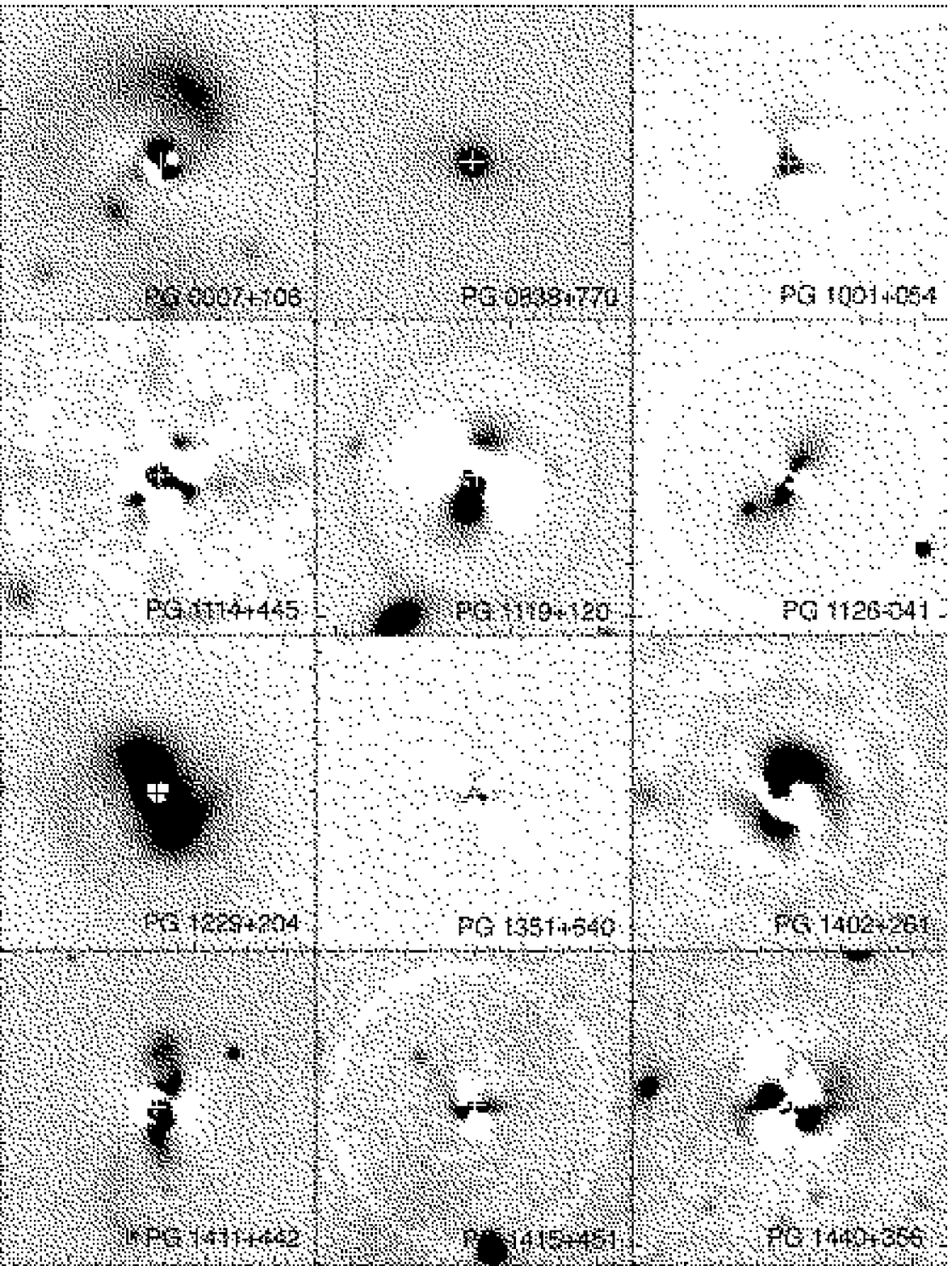}
\vspace{0.01in}
\epsscale{0.4}
\plotone{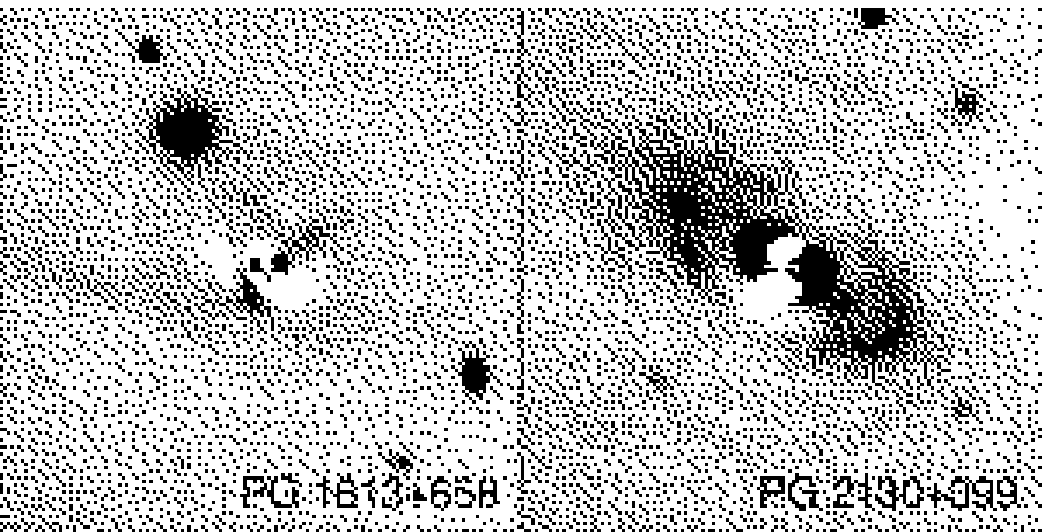}
\caption{Enhanced images of PG QSOs showing small-scale structure. All of 
the images have had a radial model of the QSO light subtracted from them. The 
location of the QSO nucleus in all cases is marked with a cross; the structure 
within a few arcseconds of this location (except in the middle panel) is 
residual error from the PSF subtraction. Tick marks are 1\arcsec.
{\bf This figure is provided separately as a JPG.}
}
\end{figure}

The nuclear subtracted image of PG 0007+106 clearly shows an arc or 
tidal tail similar to that in Mrk 231. The structure in PG 
1119+120 is clearly seen in Figure 1. The nuclear-subtracted image of PG 
1229+204 shows the galactic bar, as well as the condensations both in the bar 
and near its ends.
The data for PG 1411+442 show a north-south structure extending 
on
both sides of the nucleus. Since the northern extension 
seems to be the base of the tidal tail, it is possible the southern structure is 
actually the base of a fainter counter-tail. There are many bright 
condensations or knots in the northern tail itself. Additionally, 
there is a bright knot-like (or possibly jet-like) feature to the southwest. The 
image of PG 1613+658 shows the long tidal tail to the east and what appears to be 
a counter-tail extending west. To the northeast is the bright nuclear knot 
described by Hutchings \& Neff (1992). Additionally, there is a less luminous 
high surface brightness region to the southwest. Finally, the image of 
PG 2310+099 clearly shows the spiral structure apparent in the much 
higher resolution {\it HST} image.

\subsection{Luminosities}

Table 2 lists BIHK\p \ integrated photometry for all of the QSOs as well as 
nuclear and host galaxy luminosities and nuclear fractions.

The observed underlying host galaxies (including I Zw 1, Mrk 1014, and 
3c273) span a range in absolute magnitude 
from {\it M}$_{\rm H}$=$-$23.2 (PG 1001+054) to {\it M}$_{\rm H}$=$-$26.1 
(3c273), with a mean value of {\it M}$_{\rm H}$=$-$24.5$\pm$0.8 and a 
median of {\it M}$_{\rm H}$=$-$24.4. The H-band luminosities range from 0.5 
to 7.5 {\it L}$^*$\footnote{M$^*_H$ = -23.9 (Surace et al. 2000). 
Throughout this paper quoted values are for {\it H}$_0 = 75 \ km \ s^{-1} 
Mpc^{-1}$}
, with a median of 1.4 {\it L}$^*$ and a mean of 2.3 {\it L}$^*$. 
The cumulative distribution of the host galaxy H-band magnitudes is 
shown in Figure 7.

\begin{figure}[htbp]
\epsscale{0.7}
\plotone{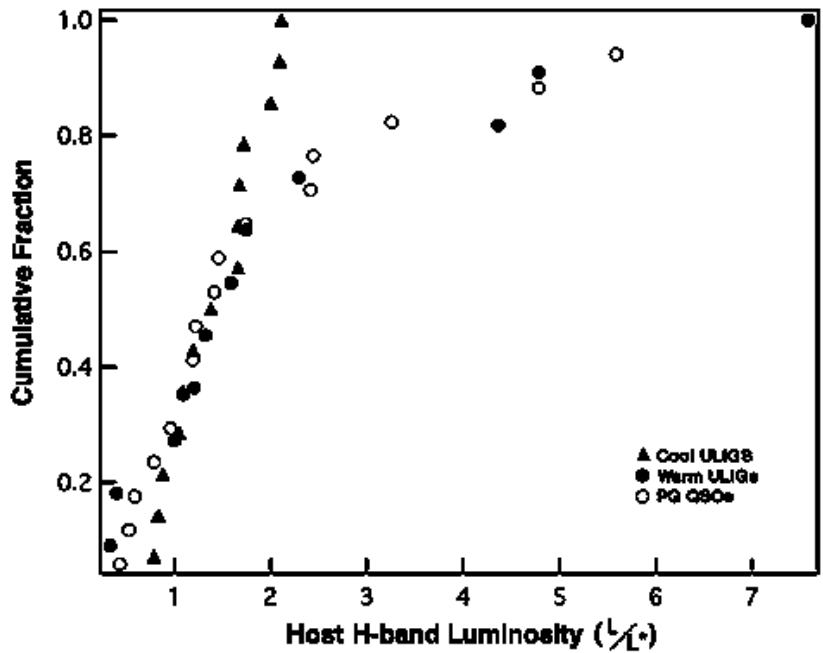}
\caption{Cumulative distribution functions of the host galaxy luminosity at 
H-band for the two ULIG samples and the far-IR excess QSO, relative to 
L$^*$. Note that the most luminous ULIG is also a QSO, hence the use 
of only one symbol.
{\bf This figure is provided separately as a JPG.}
}
\end{figure}


The fraction of the total system luminosity at a given wavelength 
originating in the QSO nucleus is 
given in Table 3. The average nuclear fraction is (BIHK\p)=(0.79, 0.67, 0.60, 
0.71)$\pm$(0.08, 0.16, 0.15, 0.10). As expected from previous studies (Sanders et 
al. 1989, McLeod \& Rieke 1995), the minimum nuclear fraction occurs near 1 
$\mu$m, which is a result of the relative SEDs of QSOs and galaxies. There is a 
considerable range in nuclear fractions: the QSO nucleus at {\it H}-band 
accounts for anywhere from 36\% to 85\% of the total H-band luminosity. On average, 
though, the QSO nucleus and the underlying host galaxy are roughly similar in 
total magnitude at {\it H}. At shorter wavelengths such as {\it B}, the QSO 
nucleus is roughly 4 times more luminous than the host.

\subsection{Colors}

Optical/near-infrared colors were derived for the QSO nuclei. Figures 
8 and 9 
illustrate the ({\it B$-$I, I$-$H, H$-$K}\p ) colors of the QSO nuclei, using the 
two rotations of the optical/near-infrared color cube introduced in Papers II \& III. 
The figures depict in this three-color basis the colors of an instantaneous 
starburst taken from Bruzual \& Charlot (1993; BC95), as well as the effects of a 
variety of reddening and emission mechanisms: thermal dust emission, a mixed 
dust and star distribution, and free-free emission. Also shown are the colors of 
a synthetic optical QSO developed by Surace et al. (1999) based on the 
properties of the PG sample as a whole, and the effects of reddening it via 
inclusion of a hot thermal dust emission component. The colors of ``warm'' 
ULIGs are shown with an  open circle.
The mean QSO nuclear colors are ({\it B$-$I,I$-$H,H$-$K}\p) = (0.99, 1.56, 
0.96)$\pm$(0.23, 0.52, 0.17). 

Table 4 presents photometry derived for the star-forming knots and other 
structures seen in some of the QSO hosts. As in Papers I, II, \& III, the 
photometry was compared to the stellar synthesis models of BC95. A model for 
the stellar (photosphere) colors from an instantaneous starburst with upper 
and lower mass cutoffs of 125 and 0.1\msun, respectively, was used. The ages 
derived from the colors are difficult to constrain, however, primarily due to a 
lack of sensitivity in the near-infrared - the detection limits are too high. In 
most cases the constraints that can be made indicate young ages for most of the 
knots ( less than $\approx$10$^8$ years). The arc in PG 0007+106, while 
apparently bluer than it's surroundings, is unconstrained due to a lack of a 
clear near-infrared detection. The nuclear knot in PG 1119+120 is very red (H-
K = 1.25), with a considerable K\p excess compared to stellar colors (H-K=0.2) 
The knot in the spiral structure of PG 1119+120 appears quite blue and the 
upper limit at H implies an age of less than 10 Myrs. Generally, the upper 
limits at H can constrain the stellar population to young ages below 10-100 
Myrs (Surace et al. 2000a) Similarly, the knots in PG 1229+204 are similarly 
blue and have ages of less than 100 Myrs old. The knots in PG 1411+442 are ill-
constrained due to large measured uncertainties, but they are also less than 
100 Myrs. The nuclear knot (3) has colors similar to many of the star-forming 
knots seen in the ULIGs (Surace et al. 2000). Finally, the nuclear knot in PG 
1613+658 is most prominent at {\it H}-band. While actually quite red optically 
(B-I$>$3.1), its near-infrared colors are typical of stellar colors (H-K\p=0.24).

The colors of the host galaxies themselves as determined from the global and 
nuclear photometry are poorly determined. This is primarily because the 
nuclei dominate the luminosity, particularly at very long ($>$ 
2$\micron$)  and 
short ($<$ 0.5 $\micron$) wavelengths. Since the average QSO nucleus is anywhere 
from 1.5 to 4 times more luminous than the host at B, relatively small errors in 
the determination of the QSO nuclear luminosity result in disproportionately 
large uncertainties in the QSO host luminosity. The mean colors of the QSO 
sample host galaxies (excluding the 3 ULIGs which are also QSOs) so derived are 
({\it B$-$I,I$-$H,H$-$K}\p) = (1.66, 1.86, 0.65)$\pm$(0.79, 0.65, 0.36) and are 
marked with a large solid circle in Figure 8 \& 9.

\begin{figure}[htbp]
\epsscale{0.55}
\plotone{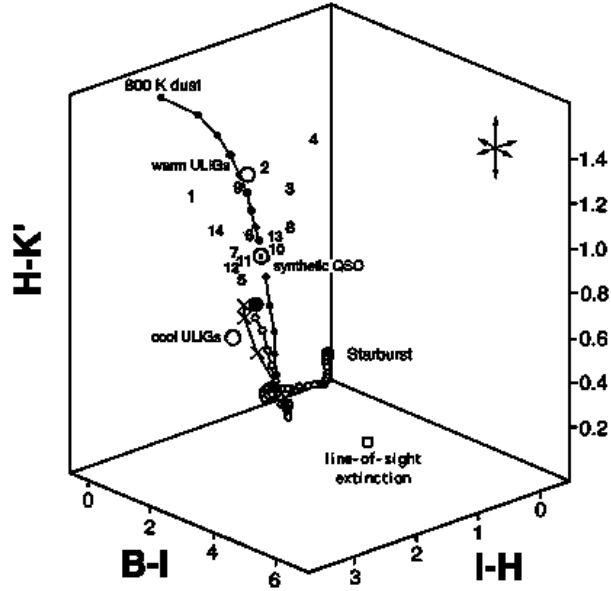}
\figcaption{\small({\it B$-$I},{\it I$-$H},{\it H$-$K\p}) color cube illustrating the colors of the QSO nuclei. 
For clarity, the QSO nuclei have been marked with numbers: (1) PG 0007+106 (2) PG 0838+770 (3) PG 1001+054 
(4) PG 1114+445 (5) PG 1119+120 (6) PG 1126-041 (7) PG 1229+204 (8) PG 1351+640 (9) PG 1402+261 (10) PG 1411+442 
(11) PG 1415+451 (12) PG 1440+356 (13) PG 1613+658 and (14) PG 2130+099. The mean host galaxy colors are 
given by the large filled circle. The cube is rotated to be orthogonal to the reddening vector, which is 
depicted by the closed boxes and represents line-of-sight extinction, i.e. a simple foreground dust screen, in 
units of {\it A}$_{\rm V}$ = 1 magnitude. 
It is derived from Rieke \& Lebofsky (1985). 
The median colors of the ``warm'' ULIGs are marked with a large circle. 
The open circles are the colors of an instantaneous starburst with 
a Salpeter IMF and mass range 0.1--125\msun and aging from 0 to 15 Gyrs, based 
on the spectral synthesis models of Bruzual \& Charlot (1993; an updated version 
called BC95 is used here).
The large dotted open circle is a synthetic QSO spectrum based on multiwavelength 
observations of all Palomar-Green QSOs and is discussed in detail in Paper II. It is 
indicative of the colors of typical optically selected quasars.
The closed circles illustrate an 800K thermal dust contribution to the colors of the 
optically selected QSOs and to a 100 Myr-old starburst. 
The joined, open circles show the effects of adding free-free emission with a 20,000 K 
electron temperature in increments of 
20\% of the total flux at K\p to the starburst.
The two sets of filled, joined circles illustrate emission from 800 K dust contributing in 
increments of 10\% to the total flux at K\p.
Finally, emission from uniformly mixed stars and dust, in units of {\it A}$_{\rm V}$=10, 30, 
and 50 magnitudes, are shown by the $\times$ symbol.
One $\sigma$ error bars are shown at upper right. Note that the line-of-sight dust extinction 
and thermal dust emission curves are nearly 
orthogonal. The QSO nuclei are very similar to the synthetic QSO colors and ``warm'' ULIGs.
}
\end{figure}

\begin{figure}[htbp]
\epsscale{0.7}
\plotone{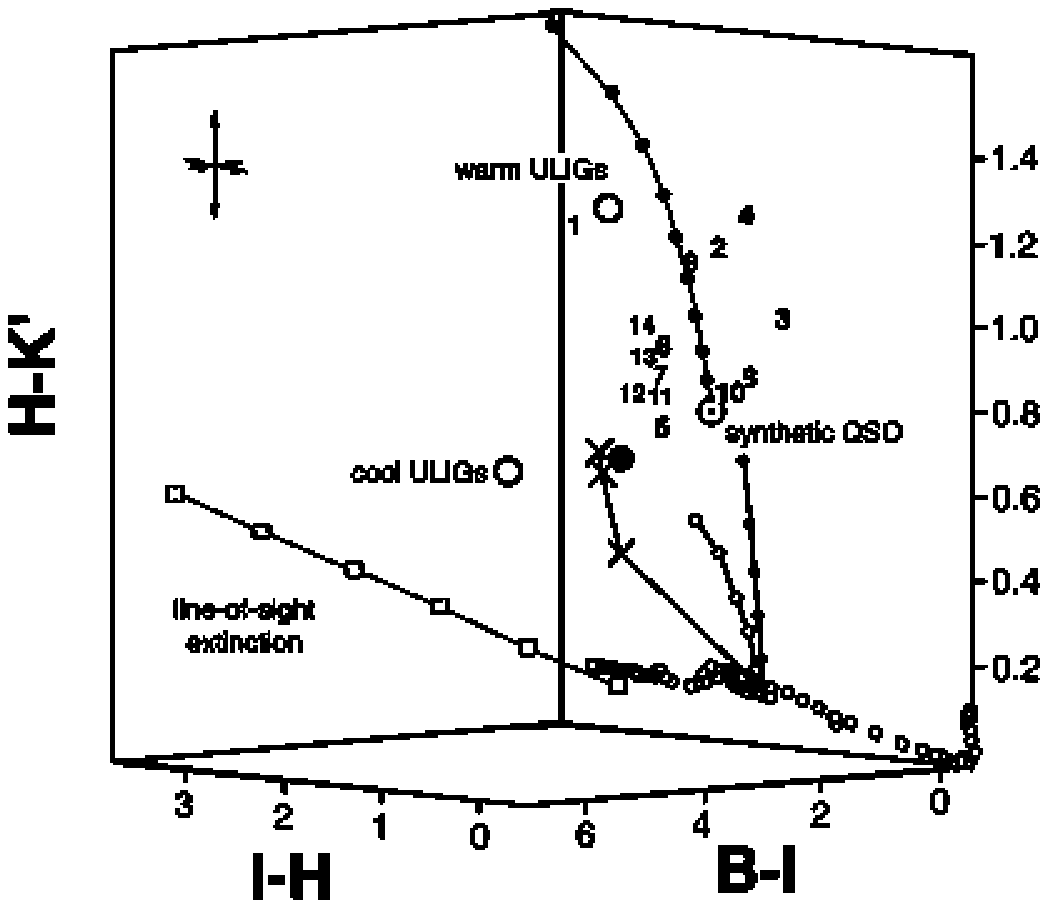}
\caption{Same as Figure 8, but rotated as to be parallel to the reddening 
vector. Most of the QSO nuclei have colors similar to the synthetic QSO colors, 
albeit with high values of ({\it H$-$K}\p). This near-infrared excess, however, 
is less than that of the warm ULIGs.}
\end{figure}

\section{Discussion}

\subsection{Morphology}

The host morphologies are 
surprisingly diverse.
Adopting the criteria for 
radio ``loudness'' of P$_{5 GHz} > 10^{24.7}$ watts Hz$^{-1}$ (Woltjer 1990) and 
the radio fluxes of Kellerman et al. (1994), we find that all of our sample QSOs 
are radio-quiet except for 3c273. Although PG 0007+106 also meets this 
definition it has alternately been classified as both radio-quiet and radio-
intermediate, and appears to have a high level of variability (Falcke 
et al. 1996, Dunlop et al. 1993).
 The one bona fide
radio-loud object, 3c273, is known to have an elliptical host.
Fully 50\% of the radio-quiet 
QSOs in our sample occur in spiral-like galaxies and at least 75\% 
occur in clearly non-elliptical hosts, which supports earlier claims 
that radio-quiet QSOs occur most often in spirals and contradicts some recent 
claims that elliptical hosts are more predominant (McLure et al. 1999).  
However, this 
result may be a selection effect: our QSO sample was 
specifically chosen to be infrared biased. It is known that infrared-selected 
samples of galaxies are biased strongly against ellipticals; they are vastly 
under-represented in {\it IRAS} surveys, presumably due to a lack of dust 
(Sulentic 1988, Surace 1998). Furthermore, this may also result from morphological type 
being a function of luminosity (Dunlop 2001). McLure et al. (1999) find that 
nearly all of the 
QSOs they observed lie in elliptical-type hosts. However, those QSOs had M$_R <$ 
-23.1 (adjusted for cosmology), which is more luminous than the majority of 
our sample (where we have estimated M$_R$ by interpolation between B 
and H-band). The low redshift of the IR-excess sample results in a bias 
towards lower luminosity QSOs, which form a continuous luminosity 
distribution with Seyfert galaxies. Our fraction of morphological types is extremely 
similar to that found for Seyfert galaxies. MacKenty (1990) found that half of 
all Seyferts were in spiral galaxies, and that half of these are barred. A similar 
result was found by McLeod \& Rieke (1995).

Four of the eight spiral hosts are clearly barred, as 
is a fifth object, also a spiral, which is
slightly too faint to be in the sample. 
The thick bars may have been detected previously and been thought to be elliptical 
hosts; deeper imaging reveals spiral arms and rings attached to them. The 
fraction of barred systems (50\% 4/8) is higher (by roughly a factor of 
2$\times$) than that usually found for intermediate-type field spirals and 
early-type spirals found in the field or in groups (Elmegreen et al. 
1990), and is similar to the result found by Hutchings \& Neff (1992). 
However, other observations in the near-infrared (Eskridge et al. 2000) have 
found a much higher fraction of bars ($\approx$ 60\%) which were hidden at 
optical wavelengths. This would indicate that the QSO host galaxy bar 
fraction is typical of the spiral galaxy population.
However, their mean peak ellipticity is fairly low (0.36), and for each system 
the typical ellipticity of the bar is somewhat lower (0.2---0.4), 
making these ``weak'' bars. This is consistent with other 
studies (Shlosman et al. 2000) which have found that Seyfert galaxies have a higher weak bar 
fraction than non-active galaxies.

Many authors have invoked galaxy mergers as a possible mechanism for 
fueling QSOs, and four of the IR-excess QSOs occur in a major merger 
system. This is similar to the 
$\approx$15\% interaction fraction found by Bahcall et al (1997).
None of the barred host galaxies is also one of the four merger systems with obvious 
tidal tails. It is possible, however, that the barred systems also 
result from mergers. Laine 
\& Heller (1999) have shown that the merger of a spiral galaxy and a much 
smaller companion can produce a barred morphology similar to that seen in 
the IR-excess PG QSOs. However, these ``minor'' mergers are qualitatively 
different from the ``major'' mergers implicated in ULIGs which involve similarly massive galaxies. 
Counting the 4 barred spirals and the 4 merger systems, interactions of some 
kind may be implicated in as many as 47\% of the QSOs. While this 
merger fraction is high, it does not approach the value of nearly 
100\% observed in ULIGs. 

Four of the galaxies with star-forming knots are also those 
that show evidence for merger activity; the fifth is also postulated 
to be a merger remnant (Hutchings \& Neff 1992). This is consistent 
with the appearance of massive star formation in mergers in general, 
and specifically with those 
found in ULIGs. These features are not as numerous as those found in 
ULIGs, despite the images being of similar sensitivity. Again, this is consistent 
with the notion of fading in the 
stellar knots.

It is worth noting that eight of these IR-excess QSOs have been 
detected in the millimeter CO (1-0) emission line (e.g. Evans et al. 
2001), indicating that their host galaxies are rich in molecular gas. 
This further strengthens the connection with the ULIGs, which are all 
known to be molecular gas rich. Two of the QSOs (PG 1351+640 and PG 
1415+451), for which no host galaxy classification could be made, are 
rich in molecular gas  and are thus likely to be located in spiral 
galaxies or on-going mergers.

Finally, it is also worth noting that there is a strong morphological selection bias built into 
the Palomar-Green Survey. In particular, the survey states that a QSO `` 
...should have a dominant starlike appearance on blue prints of the 48 inch 
(1.2 m) Schmidt Sky Atlas...'' (Schmidt \& Green 1983). This criterion was then 
used to distinguish between QSOs and Seyfert galaxies. Objects which might 
have shown broad lines but had particularly luminous host galaxies were thus 
discarded as Seyfert 1s. This morphological criterion was then further used to 
create the {\it M}$_{\rm B} < $-$23$ luminosity criterion used to define QSOs, as 
most of the objects below this criterion showed evidence for nebulosity, but 
those above did not. However, this criterion is flawed; numerous studies have 
shown that typically the QSO nucleus and its host galaxy are of comparable 
luminosity (McLeod et al. 1994ab), although this is wavelength dependent. 
Therefore, distant objects measured through large apertures might meet the 
luminosity criterion due to the presence of their host galaxies. Additionally, 
lower redshift objects whose nuclei might meet the luminosity criterion but 
whose host galaxies are also very bright and/or extended will be rejected. In 
short, the PG BQS is biased against QSOs with luminous extended hosts at low 
redshifts.
This morphological bias is likely to skew the results of this study towards 
finding systems with low luminosity hosts and towards hosts with few 
recognizable features. Other studies (Goldschmidt et al. 1992, Kohler et al. 1997) 
have also indicated quantitatively that the BQS is incomplete and heavily 
biased. In particular, it seems to be incomplete by at least a factor of 2 for low 
redshift, low luminosity QSOs ({\it z}$<$ 0.3; $-$23 $> {\it M}_{\rm B} >$ $-$24.1). 
Any results found here regarding the fractions of QSOs with luminous, 
extended (and possibly with interacting morphologies) hosts are thus likely to 
be underestimates of the true fraction. On the other hand, Kohler et al. (1997) 
also finds that the space density of high luminosity QSOs has been severely 
underestimated in the BQS. Finally, under the unified model of AGN 
(which attributes the different species of AGN to different viewing 
angles towards a central engine) there should exist 
``Type II'' QSOs whose geometry is such that the QSO nucleus is hidden from 
view by the obscuring dust torus, and evidence has been found for a sizable 
population of red QSOs entirely missed by optical QSO surveys (Webster et al. 
1995). The results presented here, particularly in regards to what fraction of 
the QSO population resides in what kind of host galaxy, should therefore be interpreted 
strictly in the light of our sample selection criteria. The question of their 
relevance to the general QSO population, whatever that might be, may have to 
wait until a better understanding of the space density of physical objects 
called ``QSOs'' is achieved.

\subsection{Luminosities}

Our H-band luminosities are  almost identical to the result found by McLeod and 
Rieke (1994a,b), who 
found a mean {\it M}$_{\rm H}$=$-$24.3 for all PG QSOs with {\it z} $<$ 0.3. 
McLeod \& Rieke split their data into two samples, a high luminosity sample 
({\it M}$_{\rm total\ B} <-$23.2; Mcleod \& Rieke 1994b) and a low luminosity 
sample ( {\it M}$_{\rm total\ B} >-$23.2; Mcleod \& Rieke 1994a). These samples 
had mean H-band luminosities of {\it M}$_{\rm H}$=$-$24.7 and $-$24, 
respectively, and they postulated that this implied that high luminosity QSOs 
were found in high luminosity hosts (McLeod \& Rieke 1994b). It was 
fair for them to compare the H-band nuclear luminosity to the B-band total integrated 
luminosity since the B-band luminosity is dominated by the QSO nucleus (see 
below) and to first order may be taken as indicative of the QSO nuclear 
luminosity. Unsurprisingly, our data show a strong correlation between B-
band nuclear and total luminosity. We do not, however, actually see a 
correlation between {\it M}$_{\rm B}$ and {\it M}$_{\rm H}$ in our data, nor 
is there is any significant correlation between the H-band luminosity of the 
host galaxy and either the total or nuclear H-band luminosity. Similarly, there 
is no statistically significant correlation between nuclear and host luminosity 
in B or I-band. However, we note our small sample size vs. that of 
McLeod et al. (1995b; 14 vs. 50) which may prevent such a correlation being 
observed. Even in the data of McLeod et al. (1995b), the effect is sufficiently 
weak that it requires binning all of the luminosities into just two groups to be 
detected. There {\it is} a statistically significant correlation between {\it K}-
band luminosities of the host and nuclei at better than the 0.995 level. This 
may support the notion that more luminous QSOs are found in more luminous 
host galaxies. This correlation may break down at short wavelengths due to 
contamination of the host galaxy light by emission from star 
formation, contamination by the QSO nucleus due to the increasingly poor 
spatial resolution at short wavelengths, and extinction by dust which is 
greater at shorter wavelengths. A similar, albeit weaker, result was found by 
Taylor et al. (1996).

The QSO host galaxies have a median H-band luminosity similar to that of 
ULIGs (Paper II, III), but have a mean luminosity and span a total range more 
like that of warm ULIGs (Paper II). Figure 7 shows the H-band host 
luminosity cumulative distribution functions (CDFs) of the three samples. All of the light attributable 
to QSO nuclei, putative AGN, and clustered star formation has been 
subtracted from the total integrated emission; the residual should 
only be emission from old starlight. As expected from their similar median 
values, all three samples have nearly identical distribution functions below 
the 80\% number fraction. Both the infrared-excess QSOs and the warm ULIGs 
have identical distribution functions, suggesting that the same underlying 
population of galaxy luminosities and hence galaxies make up the hosts. While 
to some extent this is due to the overlap (16\% of QSOs, 25\% warm ULIGs) 
between the two samples, removal of the overlap objects leaves the two CDFs 
similar. As was noted in Paper III, the cool ULIG sample seems to 
have a dearth of the very highest luminosity hosts, and as a result it's CDF 
deviates from the other two. There are several interpretations: first, and most 
importantly, the deviation is not statistically significant. Kolomogorov-
Smirnov statistics indicate that we can reject the null hypothesis that the cool 
ULIG sample is drawn from the same underlying population as the warm ULIG 
and QSO samples at only the 50\% level, mostly due to the small number 
statistics involved. Second, the warm ULIG and infrared-excess QSO samples are 
truly complete, i.e. they contain every object meeting the selection criteria, 
while the cool ULIG sample is only {\it statistically} complete, as it was drawn from a 
much larger sample of objects meeting the selection criteria. 
As a result, it consists of only about 30\% of the entire cool ULIG population. 
While the warm ULIG and QSO samples necessarily contain any existing 
systems with high luminosity hosts, there is a small likelihood that the cool 
ULIG subsample will fail to contain any such systems. 
On the other hand, the cool sample was selected randomly to have the 
same redshift distribution as the other two samples. Forcing selection 
of high redshift systems should have biased the cool galaxy CDF 
towards higher luminosity hosts, if any correlation at all exists 
between H-band luminosity and L$_{bol}$.
If this difference in CDFs is actually real, i.e., 
there is a difference in the populations of the host galaxies,
then the distributions of host luminosities are consistent with the 
infrared-excess QSOs having evolved from warm ULIGs, but that only 80\% of 
the warm ULIG/QSO populations could have evolved from an earlier cool ULIG 
state. The remaining warm ULIGs and QSOs either arose from a different 
process, by nature of their massive hosts never passed through a cool ULIG 
state, or that any such cool state was comparatively short-lived. 
Multiple evolutionary paths for ULIGs have been postulated previously 
(e.g. Farrah et al. 2001).

Of the 4 QSO hosts with direct evidence for merger activity, 3 have the highest 
H-band luminosities in the sample ({\it L}$_{\rm H} > 3.2 {\it L}^*$; PG 
0007+106, Mrk 1014, and PG 1613+658) while the fourth (PG 1411+442) is more 
nearly 1.3 {\it L}$^*$. This is consistent with their interpretation as mergers 
of two galaxies. The 5 systems that have barred morphologies that may be 
consistent with minor merger have a mean H-band luminosity of 1.2 \lstar. 

\subsection{Colors}

The derived nuclear colors are similar to those found by Elvis et 
al. (1994) for UVSX (ultraviolet-soft x-ray) QSOs and support the known
large observed scatter in QSO colors. All of the 
nuclei have ({\it H$-$K}\p) colors redder than those of both 
our modeled synthetic QSO and those derived observationally from the UVSX sample. 
This reddening is not consistent with any appreciable ($>$1 {\it A}$_{\rm V}$) 
foreground dust screen; instead, it is most consistent with varying small 
contributions (10--20\%) at {\it K}\p from hot (600--1000 K) dust, 
whose presence has 
previously been inferred in these objects (Sanders et al. 1989 and references 
therein). This reddening at K\p is less extreme than that found in the 
warm ULIG nuclei, and they are therefore found  between the regions 
in the 3-color diagram occupied by optical QSOs and the warm ULIGs. 
Since the IR-excess QSOs have colors intermediate between the ``warm'' AGN-
like ULIGs and the larger population of optically selected QSOs, possibly 
because a smaller fraction of the luminosity at {\it K}\p originates in hot dust 
and any foreground reddening screen is much thinner, this is 
consistent with 
the idea that the IR-excess QSOs are transition objects and that as they age 
much of the dust in the vicinity of the active nucleus is dissipated or destroyed. 

An alternate explanation for the near-IR excess could be that of a scattering 
dust geometry which reddens the observed QSO light. Scattering 
is greatest at shorter wavelengths, and decreases the effective extinction. However, 
an examination of Figure 7 shows that the 
reddening of the IR-excess QSOs is almost entirely due to an excess of K\p 
emission. The reddening law required to produce this would have to be nearly 
flat (wavelength-independent) at optical wavelengths. In particular, our 
observations could be reproduced if E(B-I)$\approx$ E(I-H) $\approx$ 0.2 for 
A$_{\rm V}$=1. This is a (B-I) color excess roughly four times less than that of 
the line-of-sight extinction law. This is also smaller by a factor of several than 
that derived by Calzetti (1997) for starburst galaxies, in which scattering is 
believed to be important, or modeled by Witt et al. (1992) for various realistic dust 
geometries with a central source.

The host galaxy colors are redder at K\p than expected for a stellar population 
with a Salpeter IMF. This same 
result 
was found previously by Surace et al. (1998) for the three warm ULIGs that are also PG 
QSOs. This may be a result of residual QSO light (which is much redder in the 
near-infrared than starlight) contaminating the colors of the hosts. The 
surface brightnesses were measured at the apparent half-
radius point in the host galaxies (these mean surface brightnesses appear in 
Figure 10). The colors derived 
are ({\it B$-$I,I$-$H,H$-$K}\p) = (2.16, 2.14, 0.65). This is still a considerable 
({\it H$-$K}\p) excess, considering that these values were measured far 
enough from the nucleus that residual nuclear contamination should be quite 
small. On the other hand, these colors are also very similar to those observed 
in the central regions of the cool ULIGs, and may represent a near-infrared excess 
due to recent star-formation activity (Paper III). McLeod \& Rieke (1994a) find 
a similarly high value of {\it H$-$K} when comparing their results to those of 
Dunlop et al. (1993). They note, however, that typical K-corrections to normal 
galaxy colors may result in {\it H$-$K}$\approx$0.45 (their sample being 
similar in redshift to ours). If this were combined with modest reddening ( {\it 
A}$_{\rm V}$ = 1 magnitude ) and hot dust emission, the observed colors could 
be accounted for, considering the uncertainties introduced by the 
host/nucleus decomposition.

\section{Conclusions}

\noindent 1. Host galaxies were detected in all 16 observed infrared-excess PG 
QSOs. The one QSO not observed (PG 1202+281) was already known to have an 
elliptical host.
In many cases these galaxies had observable 2-dimensional structure. At least 
47\% (8/17) have features indicative of spiral structure, 50\% (4/8) of which have 
weak
nuclear bars. Twenty-four percent (4/17) are clearly on-going major merger systems, 
and galaxy interactions may be implicated in as many as 50\% of all the QSO 
systems.

\noindent 2. The underlying host galaxies have {\it H}-band luminosities, 
which are believed to be indicative of the size of the underlying old stellar 
population, ranging from 0.5 {\it L}$^*$ to 7.5 {\it L}$^*$, with a mean 
of 2.3 
{\it L}$^*$. These luminosities are similar to those of ULIGs, although the 
``cool'' ULIGs with no clear evidence for AGN activity seem to lack the most 
massive hosts exhibited by ``warm'' ULIGs and QSOs.

\noindent 3. The QSO nuclei and their host galaxies are similar in total 
luminosity at {\it H}-band, but this ratio increases to an average of 4:1 at {\it 
B}.

\noindent 4. The QSO nuclei have near-infrared excesses relative to 
other optically selected QSO samples. This may be the result of hot 
dust emission.
The host galaxies also have considerable near-infrared excesses. This 
may be the result of recent star-formation or dust. Derived ages based on the colors of 
the observed host galaxy features are typically less than 100 Myrs.

\noindent 5. Contrary to many expectations, QSO host galaxies at low redshift 
are easiest to detect at optical and not near-infrared wavelengths. This is due 
primarily to the low background emission of the sky at optical wavelengths 
coupled with the large projected size of the extended hosts.

\acknowledgments

The authors would like to thank the telescope operators, John Dvorak and Chris 
Stewart. J.A.S. was supported by NASA grant NAG 5-3370 and by the Jet 
Propulsion Laboratory, California Institute of Technology, under contract with 
NASA. DBS was supported in part by JPL contract 
no. 961566. ASE was supported by RF9736D and AST 00-80881.
This research has made use of the NASA/IPAC Extragalactic 
Database (NED) which is operated by the Jet Propulsion Laboratory, California 
Institute of Technology, under contract with the National Aeronautics and 
Space Administration.

\appendix

\section{Notes on Individual Objects}

The following are descriptive notes explaining various features of the 
individual QSO hosts. These features are most easily seen in the near-truecolor 
images of Figure 2.

\noindent{\it PG{\ts}0007+106 = Mrk 1051 = III Zw 2} --- a single tidal arm or arc extends 
22 kpc to the north. This arm has several knots of star formation in its far end. 
Another galaxy is seen directly to the south. Although there is no apparent 
connecting structure between the two, there does seem to be some extended 
low surface brightness emission surrounding all of the galaxies in the field, 
suggesting that perhaps there are multiple interactions.

\noindent{\it PG{\ts}0050+124 = I Zw 1} --- this is also a ``warm'' ULIG and is discussed in 
more detail in Papers I \& II. It has two asymmetric spiral arms, both of which 
have knots of star formation. The galaxy disk is 30 kpc in diameter.

\noindent{\it PG{\ts}0157+001 = Mrk 1014} --- this QSO is also a ``warm'' ULIG and is 
discussed in detail in Papers I \& II. The host galaxy  is dominated by a tidal 
arm extending to the NE.

\noindent{\it PG{\ts}0838+770} --- a nuclear bar 22 kpc in projected length runs 
E-W through the host galaxy. Knots are seen near each end of the bar. A single 
spiral arm extends clockwise from each end of the bar. The galaxy itself is 
elongated perpendicular to the axis of the bar, although this may be a 
projection effect.

\noindent{\it PG{\ts}1001+054} --- small and uniform, with no distinguishable features.

\noindent{\it PG{\ts}1114+445} --- the host galaxy is diffuse and uniform, but shows 
some evidence to the south of having a small, tight spiral arm, suggesting that 
this is a face-on disk galaxy.

\noindent{\it PG{\ts}1119+120} --- a very red compact emission source lies inside the 
host galaxy envelope to the NW; with high spatial resolution it appears 
tangentially elongated. The host galaxy consists of a tilted central  elliptical 
condensation which may be a 15 kpc long bar, with an apparent disk 28 kpc in 
diameter aligned perpendicular to it. There are at least 5 additional smaller 
galaxies located within a projected distance of 25\arcsec \ of the QSO.

\noindent{\it PG{\ts}1126$-$041 = Mrk 1298} --- an elongated, elliptical host 
approximately 30 kpc in diameter. There are projections from the nucleus 
parallel to the long axis of the host which suggest that this may be a bar 
structure.

\noindent{\it PG{\ts}1202+281 = GQ Comae} --- we were unable to observe this QSO due to 
the sensitivity limitations of our fast tip/tilt guider. However, Bahcall et al. 
(1997) have observed this system with {\it HST}/WFPC2 and find that the host 
galaxy appears to be a small elliptical E1 elliptical galaxy.

\noindent{\it PG{\ts}1229+204 = Mrk 771 = TON 1542} --- a nuclear bar 14 kpc in 
projected length runs NE-SW through this host. A very blue, linear chain of 
star-forming knots extends nearly perpendicular to the bar axis at one end. 
Hutchings et al. (1992, 1994, 1995) claim that this is due to tidal disruption with 
a small companion; however, it seems more likely that these knots are 
associated with the bar itself, considering that there appears to be a similar 
blue region visible on the opposite end of the bar (Figure 2). 

\noindent{\it PG{\ts}1351+640} --- no features are discernible in the ground-based 
images.

\noindent{\it PG{\ts}1402+261} --- Bahcall et al. (1997) describes this as an SBb(r) spiral 
galaxy. The {\it HST/WFPC2} images show a clear bar with very open arms 
extending from their ends, a morphology common to several other QSO hosts 
(e.g., PG 0838+770, PG 1119+120). The ground-based images are just capable of 
showing this structure.

\noindent{\it PG{\ts}1411+442} --- a tidal loop 88 kpc long extends to the north, then 
wraps around to the east and extends south. In the model-subtracted image a 
counter-tail is seen extending south and curving to the west. In the near-
infrared, a linear structure is seen extending to the SW from the nucleus 
which may be a jet. This same jet-like feature can be seen in the optical images 
after subtraction of a radial model.

\noindent{\it PG{\ts}1415+451} --- no distinguishable features.

\noindent{\it PG{\ts}1440+356 = Mrk 478} --- the galaxy nucleus is elongated NE-SW, and 
the body of the galaxy is elongated perpendicular to this, a morphology 
common to many other QSO hosts. There is some faint evidence of shells or 
arms.

\noindent{\it PG{\ts}1613+658} --- first described in detail by Hutchings \& Neff (1992). 
A very obvious tidal tail extends 88 kpc to the east. In the near-infrared a 
second bright peak appears 2.4\arcsec \ to the NW and is most easily seen at 
{\it H}. There is a companion galaxy nearby; it is unclear if it has anything to 
do with the QSO host since there is no detection of emission linking the two. 
There is at least one other small galaxy nearby (33\arcsec \ to the SW).

\noindent{\it PG{\ts}2130+099 = UGC 11763 = Mrk 1513 = II Zw 136} --- this appears to be a 
spiral galaxy 30 kpc in diameter and inclined 60\degree \ from face-on. The 
{\it HST} images reveal the spiral structure in the host which resembles two 
concentric rings in the ground-based data. There is an odd asymmetry in the 
NE part of the host, lending to its overall peculiar appearance.

\section{Detectability of QSO Hosts}

It is commonly held that near-infrared observations are optimal for detecting 
QSO hosts (McLeod et al. 1995, Dunlop et al. 1993). This is motivated by 
the comparative SEDs of spiral galaxies and QSO nuclei; the peak of the 
galaxy SED is very near the minimum of the QSO SED, and therefore H-band 
observations yield the highest possible contrast between the galaxy host and 
the QSO nucleus. This was based on the assumption that the greatest difficulty 
in detecting QSO hosts was trying to differentiate the extended, low-surface 
brightness galaxy emission from the wings of the PSF of the extremely bright 
point-like nucleus.

However, these observations show that it is easier to 
detect QSO hosts at shorter wavelengths like B and I-band. Both our 
observations and those of others (Hutchings \& Neff 1992, Hutchings \& 
Morris 1995) indicate that typical QSO hosts are similar in gross properties to 
most galaxies. At redshifts of {\it z}$ < $0.16, a typical host galaxy 30 kpc in 
diameter is likely to be 5--10\arcsec \ in radius. This is sufficiently large that 
at a stable site like Mauna Kea under good conditions (0.5\arcsec \ FWHM at 
5000 \AA \ or 0.3\arcsec \ at 1.6$\mu$m) the wings of the PSF do not extend 
measurably to such large radii. Therefore, at low redshift ({\it z} $<$ 0.4) {\it 
the limiting factor in QSO host galaxy detection is background noise} (both 
instrumental and poisson noise from background sky emission). Modern 
near-infrared and optical 
instruments are background-limited at the depths required 
for host detection. The 2 magnitude increase in host galaxy surface 
brightness between {\it I}-band to {\it H}-band cannot offset the 5 magnitude 
decrease in sky brightness under a dark sky (this increases to 8 
magnitudes between {\it B} to {\it H}). Space-based observations, 
which have smaller sky backgrounds, would not be so affected; and 
hence the increase in host luminosity at near-infrared wavelengths would 
then increase detectability. Although the decrease in spatial resolution at long 
wavelengths with instruments like NICMOS would in theory decrease 
detectability versus optical instruments like WFPC2, as noted above the very 
large spatial scales of the QSO hosts at low redshift effectively renders this 
point moot. Only at higher redshifts ({\it z} $ > $ 0.5) would this become 
important.

\begin{figure}[tbh]
\epsscale{0.7}
\plotone{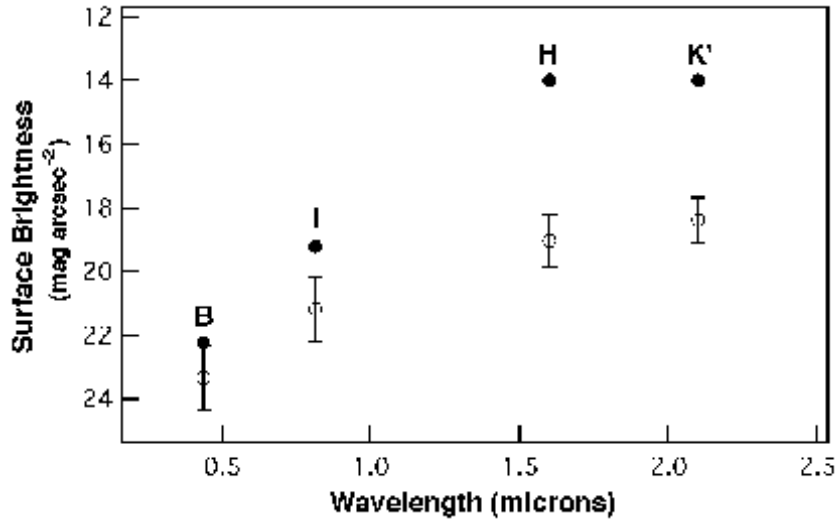}
\caption{A comparison of the surface brightnesses of the night sky (closed 
circles) on Mauna Kea (CFHT Observer's Manual 1990; Wainscoat \& Cowie 1992) 
vs. the observed mean surface brightnesses of QSO host galaxies (open circles) 
as a function of wavelength. The surface brightnesses were measured at the 
apparent half radius of the galaxies. Host galaxies are easier to detect at optical 
wavelengths such as {\it I} due to the much lower sky brightness.}
\end{figure}

Another reason for the increase in detectability lies in the relative 
morphology of galaxies at long and short wavelengths. At near-infrared 
wavelengths galaxies are relatively smooth, reflecting the distribution of old 
stars. At short wavelengths galaxies show much more complex morphologies 
due to emission from young stars and differential extinction by dust. Features 
such as spiral arms are more easily detected at short wavelengths, and 
therefore QSO host galaxies are more {\it recognizable} in filters such as {\it B} 
and {\it I}. At near-infrared wavelengths the smooth distribution of old stellar 
light in the host galaxies is more difficult to detect due to confusion with flat-
fielding errors, the wings of the QSO nuclear PSF, etc., while at short 
wavelengths the high spatial frequency features are more easily separated 
from large-scale extended background variations.


{\tiny
\begin{deluxetable}{lrrcccrrrrr}
\singlespace
\scriptsize
\tablewidth{0truein}
\tablecaption{Details of IR-Excess PG QSO Sample and Observations}
\tablehead{
\colhead{Name} &
\colhead{RA} &
\colhead{DEC} &
\colhead{z} &
\colhead{{\it L}$_{\rm ir}$/{\it L}$_{\rm blue}$\tablenotemark{a}} &
\colhead{Log {\it L}$_{\rm bol}$\tablenotemark{a}}&
\multicolumn{1}{c}{Inst.\tablenotemark{b}} &
\multicolumn{4}{c}{Exposure Times (sec)} \\[0.2ex]
\colhead{} &
\multicolumn{2}{c}{(J2000.0)} &
\colhead{} &
\colhead{} &
\colhead{{\hbox{{\it L}$_\odot$}\ }} &
\colhead{} &
\colhead{B} &
\colhead{I} &
\colhead{H} &
\colhead{K\p}}
\startdata
PG{\ts}0007+106 (III Zw 2)		& 00:10:31.0 &  10:58:29.5 & 0.089 & 0.36 & 12.23& QT & 2280 & 2520 & 1980 & 1980 \nl
PG{\ts}0050+124 (I Zw 1)\tablenotemark{c}                  & 00:53:34.9 &  12:41:36.2 & 0.061 & 0.92 & 12.32 & \nodata & \nodata & \nodata & \nodata & \nodata \nl 
PG{\ts}0157+001 (Mrk 1014)\tablenotemark{c}		& 01:59:50.1 & 00:23:41.5 & 0.163 & 1.86 & 12.68 & \nodata & \nodata & \nodata & \nodata & \nodata \nl
PG{\ts}0838+770 (VII Zw 244)		& 08:44:45.6 &  76:53:09.4 & 0.131 & 0.63 & 12.04 & QT & 1920 & 2520 & 2520 & 1800 \nl
PG{\ts}1001+054		& 10:04:20.1 &  05:13:00.5 & 0.161 & 0.49 & 12.05 & QT & 2680 & 840 & 2160 & 2040 \nl
PG{\ts}1114+445		& 11:17:06.4 &  44:13:32.6 & 0.144 & 0.98 & 12.15 & QO & 1440 & 1080 & 540 & 720 \nl
PG{\ts}1119+120 (Mrk 734)\tablenotemark{d}		& 11:21:47.1 &  11:44:18.3 & 0.050 & 1.00 & 11.48 & QT & 1080 & 1860 & 1680 & 1440  \nl
PG{\ts}1126$-$041 (Mrk 1298)	& 11:29:16.7 & $-$04:24:07.6 & 0.060 & 0.50 & 11.95 & QT & 2340 & 1860 & 900 & 1080 \nl
PG{\ts}1202+281 (GQ Comae)		& 12:04:42.2 &  27:54:11.4 & 0.165 & 1.05 & 12.11 & W\phantom{QT} & \nodata & \nodata & \nodata & \nodata \nl
PG{\ts}1226+023 (3C 273)\tablenotemark{c}			& 12:29:06.7 & 02:03:08.6 & 0.158 & 0.46 & 13.55 & \nodata & \nodata & \nodata & \nodata & \nodata \nl
PG{\ts}1229+204 (Mrk 771)		& 12:32:03.6 &  20:09:29.2 & 0.064 & 0.87 & 11.69 & QH & 1200 & 960 & 2040 & 1440 \nl
PG{\ts}1351+640		& 13:53:15.8 &  63:45:44.8 & 0.088 & 0.40 & 12.39 & QT & 2280 & 1800 & 1440 & 1560 \nl
PG{\ts}1402+261 (Ton 182)		& 14:05:16.2 &  25:55:33.7 & 0.164 & 0.49 & 12.37 & WQT & 3240 & 1800 & 2400 & 1200 \nl
PG{\ts}1411+442		& 14:13:48.4 &  44:00:13.6 & 0.090 & 0.58 & 12.00 & QO & 1440 & 1200 & 1200 & 2220 \nl
PG{\ts}1415+451		& 14:17:00.6 &  44:56:06.4 & 0.114 & 0.80 & 11.86 & QH & 1200 & 1440 & 2280 & 2400 \nl
PG{\ts}1440+356 (Mrk 478)		& 14:42:07.5 &  35:26:22.9 & 0.079 & 0.87 & 11.97 & QT & 1500 & 1680 & 1500 & 1920 \nl
PG{\ts}1613+658 (Mrk 876)		& 16:13:57.2 &  65:43:09.6 & 0.129 & 1.58 & 12.23 & QT & 900 & 720 & 2100 & 2700 \nl
PG{\ts}2130+099 (Mrk 1513)		& 21:32:27.8 &  10:08:19.5 & 0.062 & 0.50 & 11.99 & WQT & 2340 & 1260 & 2640 & 2400 \nl
\enddata
\tablenotetext{a}{Reflects new photometry presented in this work. Sample selection
was based on the photometry of Sanders et al. (1989) and Sanders et al. (1988b).}
\tablenotetext{b}{Q = UH2.2m f/31 QUIRC, T = UH2.2m f/31 Tektronix 2048, H =
UH2.2m f/31 Orbit reimaged at f/10 through HARIS spectrograph, O = UH2.2m
f/31 Orbit, W = {\it HST}/WFPC2}
\tablenotetext{c}{Also an ULIG. Data is presented in Surace et al. (1998) and Surace \& Sanders (1999).}
\tablenotetext{d}{Falls outside the QSO luminosity definition.} 
\end{deluxetable}
}



{\footnotesize
\begin{deluxetable}{lcrrrrrrrrr}
\singlespace
\scriptsize
\tablewidth{0truein}
\tablecaption{Global and Nuclear Photometry of IR-Excess PG QSOs}
\tablehead{
\colhead{Name} &
\multicolumn{1}{c}{Structure\tablenotemark{a}} &
\multicolumn{1}{c}{{\it M\tablenotemark{b}}$_{\rm B}$} &
\multicolumn{2}{c}{{\it m}$_{\rm B}$} &
\multicolumn{2}{c}{{\it m}$_{\rm I}$} &
\multicolumn{2}{c}{{\it m}$_{\rm H}$} &
\multicolumn{2}{c}{{\it m}$_{\rm K^{\prime}}$} \\[0.2ex]
\colhead{} &
\colhead{} &
\colhead{} &
\colhead{Int.\tablenotemark{c}} &
\colhead{Nuc.} &
\colhead{Int.} &
\colhead{Nuc.} &
\colhead{Int.} &
\colhead{Nuc.} &
\colhead{Int.} &
\colhead{Nuc.}}
\startdata
PG{\ts}0007+106		& TT & $-$22.0 & 15.77 & 16.2 & 14.30 & 15.1 & 11.81 
& 12.5 & 10.78 & 11.3  \nl
PG{\ts}0050+124\tablenotemark{d}& S & \nodata & \nodata & \nodata & 
\nodata & \nodata & \nodata & \nodata & \nodata & \nodata \nl
PG{\ts}0157+001\tablenotemark{d}	& TT & \nodata & \nodata & \nodata & 
\nodata & \nodata & \nodata & \nodata & \nodata & \nodata \nl
PG{\ts}0838+770		& SB & $-$22.4 &16.28 & 16.6 & 15.12 & 15.7 & 13.32 
& 14.4 & 12.66 & 13.3 \nl
PG{\ts}1001+054         & ? & $-$23.0 & 16.15 & 16.2 & 15.50 & 15.6 & 
14.48 & 14.8 & 13.41 & 13.8 \nl
PG{\ts}1114+445		& S & $-$22.8 & 16.06 & 16.2 & 14.78 & 14.9 & 13.34 
& 14.2 & 12.32 & 12.9 \nl
PG{\ts}1119+120\tablenotemark{e}		& SB & $-$21.4 & 15.09 & 15.3 & 
13.55 & 14.3 & 12.19 & 12.6 & 11.63 & 11.8 \nl
PG{\ts}1126$-$041		& S & $-$22.0 & 14.92 & 15.1 & 13.53 & 14.0 & 
11.92 & 12.3 & 11.12 & 11.3 \nl
PG{\ts}1202+281		& E & \nodata & \nodata & \nodata & \nodata & 
\nodata & \nodata & \nodata & \nodata & \nodata \nl
PG{\ts}1226+023\tablenotemark{d}		& E & \nodata & \nodata & \nodata & 
\nodata & \nodata &
\nodata & \nodata & \nodata & \nodata \nl
 
PG{\ts}1229+204		& SB & $-$21.5 & 15.55 & 16.0 & 13.96 & 15.1 & 12.33 
& 13.2 & 11.72 & 12.4  \nl
PG{\ts}1351+640         & ? & $-$23.1 & 14.63 & 14.8 & 13.65 & 13.9 & 
12.74 & 13.0 & 11.89 & 12.1 \nl
PG{\ts}1402+261		& SB & $-$23.7 & 15.51 & 15.8 & 14.98 & 15.1 & 13.22 
& 13.4 & 12.07 & 12.3 \nl
PG{\ts}1411+442		& TT & $-$23.0 & 14.79 & 15.1 & 13.88 & 14.1 & 12.42 
& 12.9 & 11.67 & 12.1 \nl
PG{\ts}1415+451         & ? & $-$22.5 & 15.90 & 16.2 & 14.70 & 15.2 & 
12.98 & 13.4 & 12.14 & 12.6 \nl
PG{\ts}1440+356		& SB & $-$22.4 & 15.15 & 15.3 & 13.77 & 14.1 & 11.83 
& 12.2 & 11.08 & 11.4 \nl
PG{\ts}1613+658		& TT & $-$23.0 & 15.67 & 16.1 & 13.98 & 14.5 & 12.19 
& 13.0 & 11.44 & 12.1 \nl
PG{\ts}2130+099		& S & $-$22.4 & 14.56 & 14.8 & 13.48 & 14.0 & 11.63 
& 12.0 & 10.78 & 11.0  \nl
\enddata
\tablenotetext{a}{ Observed host galaxy structure: TT=tidal tails, 
S=spiral structure, ?=indeterminate with no observed structure, 
E=known elliptical galaxy, B=bar.}
\tablenotetext{b}{ Uncertainties in total system measurements are 
0.05 magnitudes, and for nuclear and host galaxy measurements are 
0.12 magnitudes.}
\tablenotetext{c}{Integrated (total) and nuclear luminosity. Total 
luminosity measured within the observed I-band extent, and nuclear 
luminosity measured via point source fitting as described in the 
text.}
\tablenotetext{d}{Also an ULIG. Data is presented in Surace et al. 
(1998) and S
urace \& Sanders (1999).}
\tablenotetext{e}{Falls outside the QSO luminosity definition.}
\end{deluxetable}
}


{\small
\begin{deluxetable}{lrrrr}
\scriptsize
\singlespace
\small
\tablewidth{0truein}
\tablecaption{Nuclear Luminosity Fraction in PG QSOs}
\tablehead{
\colhead{Name} &
\colhead{B} &
\colhead{I} &
\colhead{H} &
\colhead{K\p}}
\startdata
PG{\ts}0007+106		& 0.67 & 0.47 & 0.52 & 0.61 \nl
PG{\ts}0838+770		& 0.77 & 0.57 & 0.36 & 0.58 \nl
PG{\ts}1001+054		& 0.97 & 0.93 & 0.74 & 0.70 \nl
PG{\ts}1114+445		& 0.90 & 0.88 & 0.46 & 0.58 \nl
PG{\ts}1119+120		& 0.80 & 0.49 & 0.71 & 0.83 \nl
PG{\ts}1126$-$041	& 0.82 & 0.65 & 0.39 & 0.82 \nl
PG{\ts}1202+281		& \nodata & \nodata & \nodata & \nodata \nl
PG{\ts}1229+204		& 0.64 & 0.36 & 0.44 &  \nl
PG{\ts}1351+640		& 0.84 & 0.78 & 0.79 & 0.81 \nl
PG{\ts}1402+261		& 0.78 & 0.92 & 0.85 & 0.82 \nl
PG{\ts}1411+442		& 0.78 & 0.80 & 0.65 & 0.68 \nl
PG{\ts}1415+451		& 0.73 & 0.65 & 0.69 & 0.68 \nl
PG{\ts}1440+356		& 0.88 & 0.71 & 0.72 & 0.77 \nl
PG{\ts}1613+658		& 0.70 & 0.60 & 0.46 & 0.53 \nl
PG{\ts}2130+099		& 0.83 & 0.61 & 0.72 & 0.79 \nl
\enddata
\end{deluxetable}
}

{\small
\begin{deluxetable}{lrrcrrrr}
\singlespace
\scriptsize
\tablewidth{0truein}
\tablecaption{Details of PG QSO Small Structure}
\tablehead{
\colhead{Name} &
\colhead{$\Delta$ RA\tablenotemark{a}}&
\colhead{$\Delta$ Dec} &
\colhead{Aperture\tablenotemark{b}} &
\colhead{{\it m}$_{\rm B}$}&
\colhead{{\it m}$_{\rm I}$}&
\colhead{{\it m}$_{\rm H}$}&
\colhead{{\it m}$_{\rm K^{\prime}}$} \\[0.2ex]
\colhead{} &
\multicolumn{3}{c}{arcseconds} &
\colhead{} &
\colhead{} &
\colhead{} &
\colhead{}}
\startdata
PG{\ts}0007+106	arc & $-$2.8 & 6.4 & 5.0 & 20.80 & 18.37 & $>$15.80 & $>$17.60  \nl
PG{\ts}1119+120 knot 1 & $-$0.7 & 2.8 & 1.0 & $>$19.10 & 16.93 & 16.64 & 15.39 \nl
PG{\ts}1119+120 knot 2 & 9.5 & $-$1.0 & 1.0 & 22.30 & 21.25 & $>$20.70 & $>$20.70 \nl
PG{\ts}1229+240 SW knots & -4.1 & -4.1 & 0.9 & 20.78 & 19.37 & $>$18.80 & $>$18.80 \nl
PG{\ts}1411+442 knot 1 & 3.5 & $-$18.6 & 1.0 & 23.53 & 20.70 & $>$20.70 & $>$20.20 \nl
PG{\ts}1411+442 knot 2 & $-$0.2 & 8.3 & 1.0 & 22.07 & 20.13 & $>$20.00 & $>$20.10 \nl
PG{\ts}1411+442 knot 3 & $-$1.1 & $-$1.7 & 1.0 & 20.18 & 18.91 & 16.35 & 15.88 \nl 
PG{\ts}1613+658 knot & $-$2.4 & 0.7 & 1.0 & $>$22.80 & 19.74 & 15.95 & 15.71 \nl
\enddata
\tablenotetext{a}{ offsets given relative to the QSO nucleus.}
\tablenotetext{b}{ aperture radius in arcseconds.}
\end{deluxetable}
}


\clearpage

\begin{references}

\reference{} Bahcall, J.N., Kirhakos, S., Saxe, D.H., \& Schneider, D.P. 1997, \apj, 
479, 642
\reference{} Bahcall, J.N., Kirhakos, S., \& Schneider, D.P. 1995, \apj, 450, 486
\reference{} Barnes, J. \& Hernquist, L. 1996, \apj, 471, 115
\reference{} Bessell, M.S. 1979, \pasp, 91, 589
\reference{} Canada-France-Hawaii Telescope Observer's Manual, 1990, 5-2
\reference{} Dunlop, J.S., Taylor, G.L., Hughes, D.H., \& Robson, E.I., 1993, 
\mnras, 264, 455
\reference{} Dunlop, J.S., 2001, to appear in the proceedings of "QSO Hosts 
and their Environments", Granada, January, 2001
\reference{} Elias, J.H., Frogel, J.A., Matthews, K., \& Neugebauer, 
G., 1982, \aj, 87, 1029
\reference{} Elmegreen, D.M., Elmegreen, B.G., \& Bellin, A. 1990, 364, 415
\reference{} Elvis, M.\ et al.\ 1994, \apjs, 95, 1 
\reference{} Evans, A.S., Frayer, D.T., Surace, J.A., \& Sanders, 
D.B., 2001, \aj, 119, 536
\reference{} Eskridge, P.\ B., Frogel, J.A., Pogge, R.W. et al.\ 2000, \aj, 119, 536 
\reference{} Falcke, H., Sherwood, W., \& Patnaik, A., 1996, \apj, 
471, 106
\reference{Farrah} Farrah, D., Rowan-Robinson, M., Oliver, S., ~et al.\ 2001, \mnras, 326, 1333 
\reference{} Goldschmidt, P., Miller, L. La Franca, F., \& Cristiani, S., 1992, 
\mnras, 256, 65P
\reference{} Heckman, T., Armus, L., \& Miley, G, 1990, \apjs, 74, 833
\reference{} Hodapp, K.W., Hora, J.L., Hall, D.N., Cowie, L.L. et al. 1996, New 
Astronomy, 1, 176
\reference{hooper} Hooper, E., Wilkes, B., McLeod, K., Elvis, M., Impey, C.., 
Lonsdale, C., Malkan, M. and McDowell, J. 1999, ESA SP-427, 427,893
\reference{} Hutchings, J.B., Holtzman, J., Sparks, W.B., Morris, S.C., 
et al. 1994, \apj, 
429, L1
\reference{} Hutchings, J.B. \& Morris, S.C. 1995, \aj, 109, 1541
\reference{} Hutchings, J.B. \& Neff, S.G. 1992, \aj, 104, 1
\reference{} Kellerman, K.I., Sramek, R.A., Schmidt, M., Green, R.F., \& 
Shaffer, D.B. 1994, \aj, 108, 1163
\reference{} Knapen, J.H., Shlosman, I., \& Pletier, R.F., 2000, \apj, 
529, 93 
\reference{} Kohler, T., Groote, D., Reimers, D., \& Wisotzki, L. 1997, \aa, 325, 
502
\reference{} Laine, S. \& Heller, C.H. 1999, \mnras, 308, 557
\reference{} Landolt, A., 1983, \aj, 88, 439
\reference{} Landolt, A., 1992, \aj, 104,340
\reference{} MacKenty, J.W., 1990, \apjs, 72, 231
\reference{} McLeod, K.K. \& Rieke, G.H. 1994a, \apj, 420, 58
\reference{} McLeod, K.K. \& Rieke, G.H. 1994b, \apj, 431, 137
\reference{} McLeod, K.K. \& Rieke, G.H. 1995, \apj, 454, L77
\reference{} McLure, R.J., Kukula, M.J., Dunlop, J.S., Baum, S.A., O'Dea, C.P., \& 
Hughes, D.H., 1999, \mnras, 308, 377
\reference{} Murphy, T., Armus, L., Matthews, K., et al. 1996, \aj, 
111,1025
\reference{} Neugebauer, G., Green, R.F., Matthews, K., Schmidt, M., Soifer, B.T., 
\& Bennett, J. 1987, \apjs, 63, 615
\reference{} Neugebauer, G., \& Matthews, K. 1999, \aj, 118, 35
\reference{} Perault, M., 1987, Ph.D. thesis, Univ. Paris
\reference{} Rowan-Robinson, M. 1995, \mnras, 272,737
\reference{} Sanders , D.B., Phinney, E.S.,,Neugebauer, G., Soifer, B.T., \& 
Matthews, K. 1989, \apj, 347, 29
\reference{sanders-bgs} Sanders, D.B., Soifer, B.T., Elias, J.H., Madore, B.F., 
Matthews, K., Neugebauer, G., \& Scoville, N.Z. 1988a, \apj, 325, 74
\reference{} Sanders, D.\ B., Soifer, B.\ T., Elias, J.\ H., Neugebauer, 
G., \& Matthews, K.\ 1988b, \apjl, 328, L35 
\reference{} Schmidt, M, \& Green, R.F. 1983, \apj, 269, 352
\reference{} Scoville, N.\ Z.\ et al.\ 2000, \aj, 119, 991 
\reference{} Shlosman, I., Peletier, R.\ F., \& Knapen, J.\ H.\ 2000, \apjl, 535, L83 
\reference{} Soifer, B.\ T.\ et al.\ 2000, \aj, 119, 509 
\reference {} Stockton, A., \& MacKenty, J.W., 1987, \apj, 316, 584
\reference{sulentic} Sulentic, J. 1989, \aj, 98, 2066
\reference{} Surace, J.A., Sanders, D.B., Vacca, W.D., Veilleux, S., \& Mazzarella, 
J.M., 1998, \apj, 492, 116 (Paper I)
\reference{} Surace, J.A., \& Sanders, D.B., 1999, \apj, 512, 162 (Paper II)
\reference{} Surace, J.A.,  Sanders, D.B., \& Evans, A.S., 2000a, \apj, 529, 170 
(Paper III)
\reference{} Surace, J.A., Sanders, D.B., 2000b, \aj, 120, 604 (Paper IV)
\reference{} Taylor, G.L., Dunlop, J.S., Hughes, D.H., \& Robson, E.I., 1996, 
\mnras, 283, 930
\reference{toomre}  Toomre, A., \& Toomre, J. 1972, \apj, 178, 623
\reference{} Veilleux, S., Kim, D.-C., \& Sanders, D. B. 1999, ApJ, 522, 
113
\reference{} Wainscoat, R.J. \& Cowie, L.L. 1992, \aj, 103, 332
\reference{} Webster, R.L., Francis, P.J., Peterson, B.A., Drinkwater, M.J., \& 
Masci, F.J., 1995, \nat, 375, 469
\reference{} Whitmore, B.~C.~\& Schweizer, F.\ 1995, \aj, 109, 960 
\reference{Witt} Witt, A.~N., Thronson, H.~A., \& Capuano, J.~M.\ 1992, \apj, 393, 611 
\reference{} Woltjer, L. 1990, in Active Galactic Nuclei 
(Springer-Verlag: Heidelberg), 1


\end{references}
\end{document}